\documentclass{article}
\usepackage{fullpage}
\usepackage{tikz}
\usepackage{amsmath}
\usepackage{amssymb}
\usetikzlibrary{shapes,arrows}

\definecolor{colora}{RGB}{0,100,255}
\definecolor{colorb}{RGB}{230,154,0}

\newcommand{\ave}[1]{\left \langle #1 \right \rangle}
\newcommand{\order}{\mathcal{O}}
\newcommand{\E}{\mathbb{E}}
\newcommand{\diff}[2]{\frac{\mathrm{d} #1}{\mathrm{d} #2}}
\newcommand{\floor}[1]{\left\lfloor #1 \right\rfloor}

\title{Edge-based compartmental modeling for epidemic spread Part II: Model Selection and Hierarchies}
\author{Joel C. Miller\footnote{Center for Communicable Disease Dynamics, Department of Epidemiology, Harvard School of Public Health}~\footnote{Fogarty International Center, NIH}\and Erik M. Volz\footnote{Department of Epidemiology, University of Michigan, Ann Arbor}}

\begin{document}

\maketitle

\begin{abstract}
We consider the edge-based compartmental models for epidemic spread developed in~\cite{miller:ebcm_overview}.  We show conditions under which simpler models may be substituted for more detailed models, and in so doing we define a hierarchy of epidemic models.  In particular we provide conditions under which it is appropriate to use the standard mass action SIR model, and we show what happens when these conditions fail.  Using our hierarchy, we provide a procedure leading to the choice of the appropriate model for a given population.  Our result about the convergence of models to the Mass Action model gives clear, rigorous conditions under which the Mass Action model is accurate.
\end{abstract}

\section{Introduction}

The spread of epidemics through populations is affected by many factors such as the infectiousness of the disease, the duration of infection, the distribution of contacts through the population, and the typical duration of contacts.  Typically for predicting epidemic spread or intervention effectiveness, we want an accurate, but simple, model that captures the relevant effects.  

In our earlier work~\cite{miller:ebcm_overview}, we introduced the edge-based compartmental modeling approach for the spread of SIR diseases in populations with different contact dynamics.  In particular we showed that edge-based compartmental models can capture contact duration and social heterogeneity (variation in contact levels) simultaneously in mathematically and conceptually simple terms.  The models we studied all assumed that the population was made up of individuals who were identical except for their contact levels.  We modeled the population as a network, with nodes representing individuals joined by edges representing potentially transmitting contacts.  We also assumed a simple disease, with transmission occuring at rate $\beta$ per edge and recovery occuring at rate $\gamma$.  In this paper we investigate the relationships between models and how to choose the simplest model appropriate for a given population.

The models discussed in~\cite{miller:ebcm_overview} can each be classified by two features: the first is how contact duration is determined and the second is how contact levels are distributed through the population.  Table~\ref{tab:summary}  summarizes the models and their underlying assumptions.

The contact duration falls into three possibilities: it can be permanent, finite, or fleeting.  In the permanent case, a contact that exists at any time has always existed and will always exist.  In the finite case contacts may change over time.  In the fleeting case, contacts are so brief that over any macroscopic time scale an individual samples a very large number of neighbors, and so it is safe to assume that the total contact time with infected individuals matches its expected value.

The distribution of contact levels can be split into two types.  In one set of models (expected degree models), we assign an expected degree $\kappa$ to a node.   The probability that an edge exists between two nodes is proportional to the expected degrees of each node.  Edges are created independently of one another, so the existence of an edge between $u$ and $v$ does not alter whether an edge can exist between $u$ and $w$.  Note that the expected degree can take any non-negative real value, and we assume that the distribution of $\kappa$ is given by the probability density function $\rho(\kappa)$.  In the other set of models (actual degree models) we assign an actual integer number of contacts $k$, the degree, to each individual though in the case of the dormant contact model not all of these contacts must be active at all times.  We think of an individual as having $k$ \emph{stubs} (or half-edges) which pair randomly with stubs of other nodes to form edges.  In this case the existence of an edge between $u$ and $v$ removes an available stub from $u$, and so it affects the probability of an edge between $u$ and $w$.  The distribution of degrees is given by the probability mass function $P(k)$.


The distinction between the expected degree models and the actual degree models becomes apparent when we calculate the probability that an individual is susceptible.  For the expected degree models, there is a continuum of risk levels and so we will have to calculate the per-expected degree probability of not having been infected.  In contrast for the actual degree models the risk is discretized.  The calculation is slightly different in each case, but the underlying concepts are the same.

\begin{table*}
\newcommand\T{\rule{0pt}{2.4ex}}
\newcommand\B{\rule[-1.1ex]{0pt}{0pt}}
\begin{center}
\begin{tabular}{|c|c|c|}
\hline
\T{}\textbf{Model}\B{}   &   \textbf{Contact Duration} &
\textbf{Heterogeneity type} \\\hline\hline%
\T{}Configuration Model (CM)\B{}& Permanent & Actual degree $k$\\\hline
\T{}Mixed Poisson (MP)\B{}&Permanent & Expected degree $\kappa$\\\hline
\T{}Dynamic Variable-Degree (DVD)\B{}&Finite & Expected degree $\kappa$\\\hline
\T{}Dynamic Fixed-Degree (DFD)\B{}&Finite
& Actual degree $k$\\\hline
\T{}Mean Field Social Heterogeneity (MFSH)\B{}& Fleeting & Contact rate $k$ or $\kappa$\\\hline
\T{}Dormant Contact (DC)\B{}&Finite
active and dormant & Maximum degree $k_m$\\\hline
\T{}Mass Action (MA)\B{}&Fleeting
& None\\\hline
\end{tabular}
\end{center}
\caption{The basic models and their underlying assumptions.  Contacts are assumed to form and break at some rate which can vary from zero (permanent) to infinite (fleeting).  Depending on the process governing contact formation, we may know the actual degree $k$ of an individual or we may know its expected degree.}
\label{tab:summary}
\end{table*}

The equations produced by edge-based compartmental models are surprisingly simple.  Nevertheless the cases with simpler assumptions lead to simpler equations, and so it is worth knowing what conditions allow the use of a simpler model.  In general when contact duration is short compared to the infection and recovery time scales it is safe to use the fleeting contact models, while if contact duration is long compared to the duration of the epidemic it is safe to use the permanent contact models.  However, there are some less obvious limits.  If the average degree $\ave{K}$ is large while the rate of transmission per edge scales such that $\beta$ is of order $1/\ave{K}$ and the recovery rate $\gamma$ is fixed, then the probability that any given edge transmits even once is small.  The probability it transmits twice is negligible: the disease ``sees'' an edge at most once, so whether it is permanent, fleeting, or finite has no impact on disease spread.  Thus we may treat the model as if the contacts are fleeting, which simplifies the equations.  If further the contact distribution is such that the contact levels are generally close to the mean, then we can neglect variation in contact levels, and turn to the simple mass action model of~\cite{kermack,andersonmay}.  The precise condition required for this is somewhat technical: we must have $|\ave{K^4}-\ave{K}^4|/\ave{K}^4 \ll 1$ and $\ave{K} \gg 1$ with $\hat{\beta}=\beta\ave{K}$ fixed.  When this does not hold, there can be a significant deviation away from the mass action model.

In this paper we begin by providing a flow chart leading to selection of the appropriate model for a given population.  Following the flow chart, we introduce the hierarchy of the models which underlies the flow chart.  We describe the precise assumptions of the models and demonstrate some of the simpler parts of the hierarchy.  After that, we consider some of the more difficult aspects of the hierarchy.  We finally discuss some of the implications and limitations of our approach.  In particular, we note that we give a simple heuristic for when the Mass Action equations are appropriate.  In the Appendix we provide more rigorous justifications for the claims in the hierarchy section.  Because of its importance, we include the rigorous proof of conditions under which the Mass Action equations hold in the main text.


\section{Model Selection}
\label{sec:flow}
\tikzstyle{decision} = [diamond, draw, fill=colora!20, 
    text width=4.5em, text badly centered, node distance=3cm, inner sep=0pt]
\tikzstyle{block} = [rectangle, draw, fill=colorb!20, 
    text width=5em, text centered, rounded corners, minimum height=4em]
\tikzstyle{line} = [draw, -latex']
\begin{figure}
\begin{center}
\begin{tikzpicture}[node distance = 2cm, auto]
\node [block] at (0,4) (START) {START};
\node [decision] at (3.25,4) (Q1) {Q1};
\node [decision] at (8.125,4) (Q2) {Q2};
\node [decision] at (1.25,0) (Q3) {Q3};
\node [decision] at (6.25,0) (Q4) {Q4};
\node [decision] at (10,0) (Q5) {Q5};
\node [decision] at (13.75,0) (Q6) {Q6};
\node [block] at (0,-4) (MA) {MA};
\node [block] at (2.5,-4) (MFSH) {MFSH};
\node [block] at (5,-4) (MP) {MP};
\node [block] at (7.5,-4) (CM) {CM};
\node [block] at (10,-4) (DFD) {DFD};
\node [block] at (12.5,-4) (DVD) {DVD};
\node [block] at (15,-4) (DC) {DC};
\path [line] (START) -- node {} (Q1);
\path [line] (Q1) -- node {Yes} (Q3);
\path [line] (Q1) -- node {No} (Q2);
\path [line] (Q3) -- node {Yes} (MA);
\path [line] (Q3) -- node {No} (MFSH);
\path [line] (Q2) -- node {No} (Q5);
\path [line] (Q2) -- node {Yes} (Q4);
\path [line] (Q4) -- node {Actual} (CM);
\path [line] (Q4) -- node [left] {Expected} (MP);
\path [line] (Q5) -- node {Yes} (DFD);
\path [line] (Q5) -- node {No} (Q6);
\path [line] (Q6) -- node {Yes} (DVD);
\path [line] (Q6) -- node {No} (DC);
\end{tikzpicture}
\end{center}
\begin{itemize}
\item Q1: Is the average degree $\ave{K}$ large and does the probability that an infected node recovers before transmitting scale like $1/\ave{K}$?\\ or\\
 Is an existing edge from an infected neighbor much more likely to be deleted than to have infection transmit?
\item Q2: Do almost all contacts stay the same during the time scale of the epidemic?
\item Q3: Is $|\ave{K^4}-\ave{K}^4|/\ave{K}^4 \ll 1$ where $\ave{K^4}$ is the average fourth power of the degree and $\ave{K}$ is the average degree?
\item Q4: Can we measure the actual number of contacts an individual has, or just the expected number?
\item Q5: Does a new contact only form when an existing contact breaks?
\item Q6: For a given individual, are new contacts created at a constant rate regardless of the existing number of contacts?
\end{itemize}
\caption{\textbf{Flow chart for model choice}.  Depending on the how the edge turnover rates compare to the infection and recovery rates, as well as the degree distribution, we arrive at different models.  The model abbreviations are as in Table~\ref{tab:summary}.  This flow chart assumes that all structure in the population is due to heterogeneity in contact levels.}
\label{fig:flowchart}
\end{figure}

In Figure~\ref{fig:flowchart} we present a flow chart that can be used to select the appropriate model for a given population.  The conditions depend on the degree distribution and the rate at which edges change.  These are relatively straightforward to measure for a population~\cite{mossong:PLOScontacts,wallinga:contact_survey,salathe:network}.  From the observations of population and disease parameters, we can choose the simplest model to accurately represent disease spread in a given population.  The equations for each model are described in Section~\ref{sec:hierarchy}.

Most existing modeling of infectious disease spread are based on the Mass Action model~\cite{kermack}.  This flow chart gives appropriate conditions under which this model is reasonable.  In general, we must have most degrees close to the average degree, but at the same time either contacts must have high turnover or the probability of transmission per contact is very low.  Under these conditions, it is appropriate to use the MA model.

Before using this flow chart it is always prudent to be sure that the population does not violate other assumptions of the models.  For example, different contact structure between age groups may require more consideration.  If there are important features not captured by these models it may be possible to develop a custom model that captures the relevant detail (see~\cite{miller:ebcm_structure,volz:clustered_result}).  Otherwise we may not be able to rely on edge-based compartmental models and may have to use simulation.

The questions asked are somewhat vague in the sense that whether a number is large or not is somewhat a matter of opinion.  In Section~\ref{sec:hierarchy} we address this in more detail.  We show how the results of the models converge as the parameter values change.

\section{The model hierarchy}
\label{sec:hierarchy}

In this section we investigate the hierarchy of Figure~\ref{fig:basic_hierarchy} underlying the flow chart in Figure~\ref{fig:flowchart}.  We first give a brief overview of the standard simple model, the Mass Action model.  We then consider the hierarchy of models.  We find it convenient to consider the expected degree models before the actual degree models.  We consider the models roughly in order of increasing complexity, explaining the underlying assumptions and giving the equations from~\cite{miller:ebcm_overview}.  If we can reduce the model to another model, we explain why this should happen and give some details of the mathematical explanation.  More complete derivations are in the Appendix. 

The Mean Field Social Heterogeneity models require further attention.  We have two formulations, one in terms of expected degree $\kappa$ and the other in terms of actual degree $k$.  We will not address this initially while we discuss the main structure of the hierarchy, but at the end of this section we show that the two formulations are mathematically equivalent in the sense that each can be derived from the other.  So we do not distinguish between the two models in the hierarchy of Figure~\ref{fig:basic_hierarchy}.  We will also show that all other models reduce to Mean Field Social Heterogeneity in the large $\ave{K}$ limit if $\beta\ave{K}$ and $\gamma$ are both constant.  In turn we have conditions under which the Mean Field Social Heterogeneity model reduces to the Mass Action model.  Given any model with $\beta \ave{K}$ and $\gamma$ fixed, $\ave{K} \to \infty$ and $\ave{K^4}/\ave{K}^4 \to 1$, we arrive at the Mass Action model\footnote{It is worth noting that in the \emph{expected degree} formulations the average fourth power of the degree and the average fourth power of the expected degrees are not equal, but if the ratio with $\ave{K}^4$ tends to $1$ for either, it does for the other as well.}.

\tikzstyle{int}=[draw, fill=blue!20, minimum size=2em, text width=7em]

\begin{figure}
\begin{center}
\begin{tikzpicture}[node distance=1cm,auto,>=latex']
\node [int] at (0,0) (DC) {Dormant contacts};
\node [int] at (6,-5) (DFD)  {Dynamic Fixed-Degree};
\node [int] at (-6,-5) (DVD) {Dynamic Variable-Degree};
\node [int] at (6,-12) (CM) {Configuration Model};
\node [int] at (-6,-12) (MP) {Mixed Poisson};
\node [int] at (0,-10) (MFSH) {Mean Field Social Heterogeneity};
\node [int] at (0,-16) (MA)  {Mass Action};
\path[->, thick, bend right] (DC)  edge node [left, rotate=-90, xshift = 1.1cm, yshift = -0.5cm] {\parbox{1in}{$\eta_1/\beta \to \infty$, \\ $\eta_2/\eta_1$ constant}} (MFSH);
\path[->, thick] (DC)  edge node [rotate=320] {$\eta_2/\eta_1 \to 0$} (DFD);
\path[->, thick] (DC)  edge node [left,rotate=40, xshift = 2cm, yshift=0.8cm] {\parbox{1.7in}{$P(k_m)= \int_{\frac{\eta_1k_m}{\eta_2}}^{\frac{\eta_1(k_m+1)}{\eta_2}} \rho(\kappa) \, \mathrm{d}\kappa$, \\ 
$\eta_2/\eta_1\to\infty$}} (DVD);
\path[->, thick] (DFD)  edge node {$\eta t\to 0$} (CM);
\path[->, thick] (DVD) edge node [left] {$\eta t\to 0$} (MP);
\path[->, thick, dashed] (CM)  edge node [below] {$P(k)= \int_0^\infty \frac{e^{-\kappa}\kappa^k}{k!}\rho(\kappa)\,\mathrm{d}\kappa$} (MP);
\path[->, thick] (DFD) edge node [rotate=40, yshift = 0.75cm, xshift=-0.6cm] {$\eta/\beta \to \infty$} (MFSH);
\path[->, thick] (DVD) edge node [ rotate =320, xshift = -0.5cm] {$\eta/\beta \to \infty$} (MFSH);
\path[->, thick, dashed, bend right] (MFSH) edge node [left, near end] {\parbox{1.5cm}{\begin{center} No contact rate variation
    \end{center}  }} (MA);
\path[->, thick, bend left] (DFD)  edge node[rotate = 40, xshift =-1cm,yshift=-0.1cm] {\parbox{1in}{$\ave{K} \to \infty$,\\ $\beta \ave{K}$ constant} } (MFSH);
\path[->, thick, bend right] (DVD)  edge node[rotate = 320, below, xshift = 0.2cm] {\parbox{1in}{$\ave{K} \to \infty$,\\ $\beta \ave{K}$ constant} } (MFSH);
\path[->, thick] (CM) edge  node[rotate = 345, xshift = 1.2cm] {\parbox{1in}{$\ave{K} \to \infty$,\\ $\beta \ave{K}$ constant} } (MFSH);
\path[->, thick] (MP) edge  node[rotate = 15, below, xshift = 0.6cm] {\parbox{1in}{$\ave{K} \to \infty$,\\ $\beta \ave{K}$ constant} } (MFSH);
\path[->, thick, bend left] (DC) edge  node [right, rotate = 270, xshift = -1.4cm, yshift = 0.5cm] {\parbox{1in}{$\ave{K} \to \infty$,\\ $\beta \ave{K}$ constant} } (MFSH);
\path[->, thick, bend left] (MFSH) edge node [right, near end] { \parbox{1.1in}{$\frac{\ave{K^4}}{\ave{K}^4} \to 1$,\\ $\beta \ave{K}$ constant}
 } (MA);

\end{tikzpicture}
\end{center}
\caption{The hierarchy structure.  The edges summarize the relations between the models.  A solid arrow from one model to another represents that the target model can be derived as a limiting case of the base model.  We give heuristic explanations in the text for these and more rigorous derivations in the Appendix.  A dashed arrow denotes that the target model can be derived as a special case of the base model.  The dashed arrows are straightforward and the justification is given in the main text.}
\label{fig:basic_hierarchy}
\end{figure}
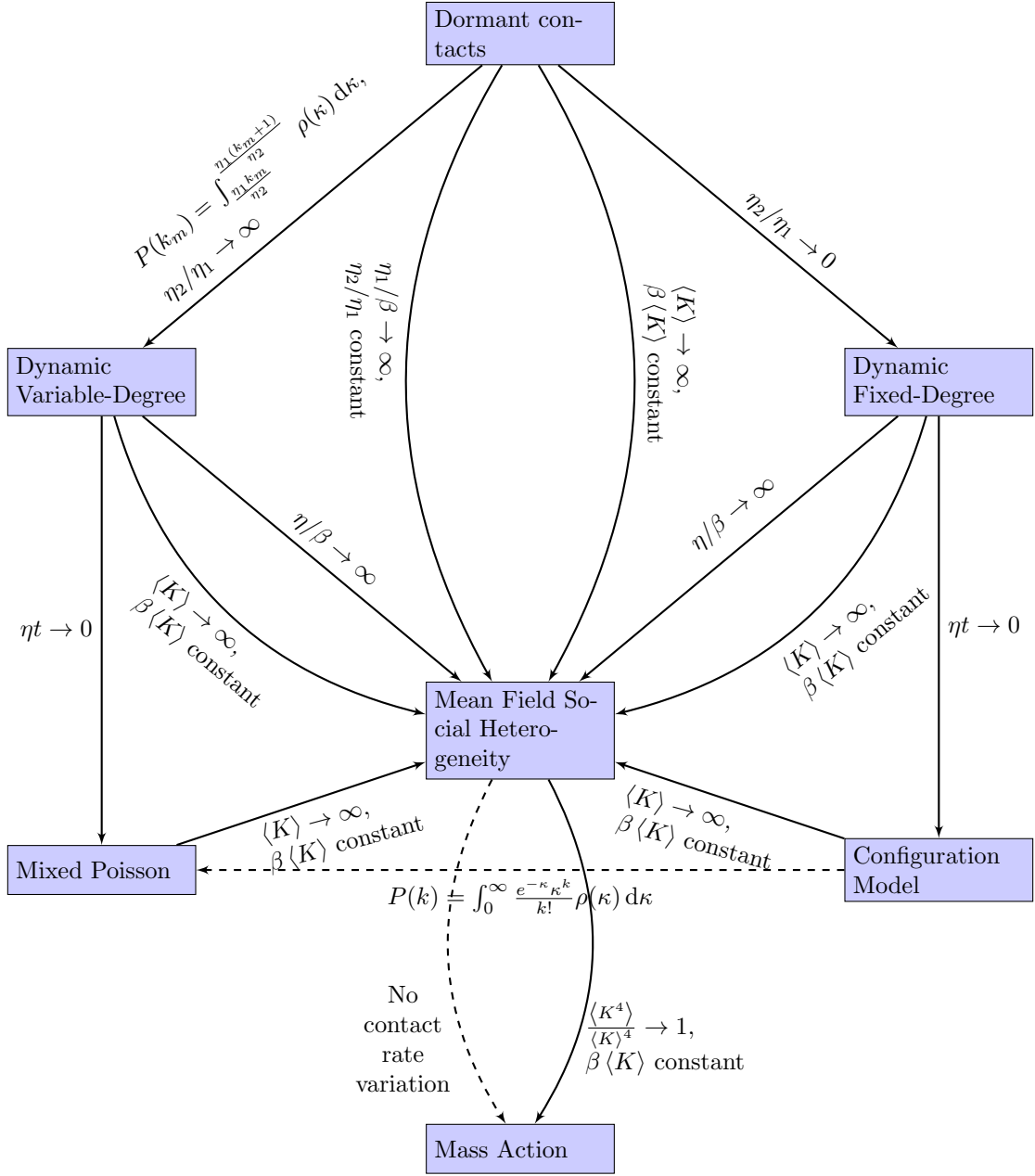

\subsubsection{The Mass Action model}
The Mass Action (MA) model assumes that all individuals have the same rate of contact formation $k$ (or $\kappa$) and contacts are sufficiently short that we may neglect repeated contacts.  The transmission rate per contact is $\beta$, and so the combined transmission rate is $\hat{\beta} = \beta k$.  If we take $S$, $I$, and $R$ to be the proportion of the population that is infected, this leads to 
\[
\dot{S} = -\hat{\beta} I S \, , \qquad\qquad \dot{I} = \hat{\beta} I S - \gamma I \, , \qquad\qquad \dot{R} = \gamma I
\] 
However, this may be simplified somewhat by noting that $S+I+R=1$.  We replace the equation for $\dot{I}$ with $I = 1-S-R$
\[
\dot{S} = -\hat{\beta} I S \, , \qquad\qquad I = 1- S - R \, , \qquad\qquad \dot{R} = \gamma I
\]

\subsection{Expected degree models}
We now study the expected degree models for which each node has an expected degree $\kappa$ assigned using the probability distribution function $\rho(\kappa)$.  We allow $\kappa$ to be a continuous variable.  At any given time, the probability that two nodes $u$ and $v$ share an edge  is proportional to $\kappa_u\kappa_v$, and each edge is assigned independently of all others.

We briefly sketch the approach used to derive equations in these networks.  Full details are in~\cite{miller:ebcm_overview}.  We define $\Theta$ as a function of time such that $1-\Theta$ represents the per-unit $\kappa$ probability of having been infected.  To be precise let $u$ and $v$ be nodes with $\kappa_v = \kappa_u+\Delta \kappa$ and $\Delta \kappa \ll 1$.  If the probability that $u$ is susceptible is $s(\kappa_u,t)$, then the probability that $v$ is still susceptible is $s(\kappa_u+\Delta\kappa,t)=[1-(1-\Theta)\Delta \kappa]s(\kappa_u,t) + \order(\Delta \kappa^2)$.  Then taking $\Delta \kappa \to 0$, we have $\partial s/\partial \kappa = (1-\Theta) s$.  So $s(\kappa,t)=\exp[-\kappa(1-\Theta)]$ and the probability a random node is susceptible is $S(t) = \Psi(\Theta(t))$ where
\[
\Psi(x) = \int_0^\infty e^{-\kappa(1-x)} \rho(\kappa) \, \mathrm{d}\kappa
\]
The difference in the various expected degree models is in how long edges last: they may be permanent, fleeting or finite.  We begin by considering permanent edges.

\subsubsection{Mixed Poisson}
In the Mixed Poisson (MP)  model, the population is assumed to be static, so that if a contact ever exists, then it has always existed and will always exist.  The governing equations are
\begin{align*}
\dot{\Theta} &= - \beta \Theta + \beta \frac{\Psi'(\Theta)}{\Psi'(1)} + \gamma(1-\Theta)\\
\dot{R} &= \gamma I \, , \qquad\qquad S = \Psi(\Theta) \, , \qquad\qquad I = 1-S-R
\end{align*}

\subsubsection{Mean Field Social Heterogeneity [expected degree formulation]}
We now consider the opposite limit where contacts are fleeting.
The Mean Field Social Heterogeneity (MFSH) model~\cite{andersonmay,may:hivdynamics, may:dynamics,moreno,pastor-satorras:scale-free} generalizes the MA model by allowing for variations in contact rate among the people.  At any given time the node is expected to have $\kappa$ contacts, but those contacts change over rapidly.  The governing equations are
\begin{align*}
\dot{\Theta} &= -\beta +\beta\frac{\Psi'(\Theta)}{\Psi'(1)} + \gamma(1-\Theta)\\
\dot{R} &= \gamma I \, , \qquad\qquad S = \Psi(\Theta) \, , \qquad\qquad I = 1-S-R
\end{align*}
Note that these equations differ in only one term from the MP equations.

The only difference in the assumptions of the MFSH model and the MA model is that the MA model assumes all contact rates are the same.  Indeed, if all expected degrees are the same $\kappa$, then $\Psi(\Theta) = \exp[-\kappa(1-\Theta)]$.  We set $R = \gamma(1-\Theta)/\beta$, \ $I=1-S-R$, and $S = \Psi(\Theta)$.  We first note that with this $\Psi(\Theta)$, we find $\dot{\Theta} = -\beta I$.  Setting $\hat{\beta} = \beta \kappa$ and taking the time derivatives of $S$ and $R$, we see that $\dot{S} = -\hat{\beta}IS$ and $\dot{R} =\gamma I$.  Thus we have arrived at the MA equations.

More generally we expect that if the variation in contact rate is sufficiently small, the MFSH model should behave like the MA model.  
In Section~\ref{sec:MFSH2SIR} we discuss this further.

\subsubsection{Dynamic Variable-Degree}
In the Dynamic Variable-Degree (DVD) model, an individual may create or delete edges at any time.  A node with expected degree $\kappa$ creates edges at rate $\kappa\eta$.  Any existing edge breaks at rate $\eta$, so a node with expected degree $\kappa$ will on average have $\kappa$ edges, though the value fluctuates.  The governing equations are
\begin{align*}
\dot{\Theta} &= -\beta\Theta + \beta\frac{\Psi'(\Theta)}{\Psi'(1)} + \gamma(1- \Theta) + \eta\left(1- \Theta - \frac{\beta}{\gamma} \Pi_R\right) \, ,\\
\dot{\Pi}_R &= \gamma \Pi_I \, ,\qquad \Pi_S = \Psi'(\Theta)/\Psi'(1) \, ,\qquad \Pi_I = 1 - \Pi_S - \Pi_R \, , \\
\dot{R} &= \gamma I \, ,\quad\qquad S = \Psi(\Theta) \, ,\quad\qquad I=1-S-R \, .  
\end{align*}
The new variables $\Pi_S$, $\Pi_I$, and $\Pi_R$ give the probabilities that a newly formed edge will connect to a susceptible, infected, or recovered node respectively.

If the changeover rate is sufficiently fast we anticipate that the model should reduce to the MFSH model.  This statment can be made precise by considering a susceptible node $u$.  The node $u$ is constantly creating new edges and breaking existing edges.  Its risk of infection depends on how many infected neighbors it has.  If an edge to an infected neighbor is likely to break before the neighbor transmits, then there is little correlation between the risk to node $u$ at different times: knowing that a neighbor has not infected $u$ yet says little about its current status and so the duration of contact becomes negligible.  In mathematical terms this means if $\eta/\beta$ is large, we anticipate that the DVD model reproduces the MFSH model.  This is made more precise in the Appendix.

In the opposite limit, we would expect that having small $\eta$ leads to an effectively static network, so the MP model should result.  In fact this is true, but the precise condition is somewhat more subtle than might be anticipated.  It is not enough that $\eta/\beta$ and $\eta/\gamma$ be very small because the epidemic can last for many generations.  In practice the static model will work well at early times, but may fail at later times as the contact structure accumulates changes.  We set $t_0$ to be a time early enough that the number of infections by time $t_0$ is very small, but large enough that shortly thereafter the total number of infections is no longer negligible.  At some later time $t$ the MP model will be reasonable if $\eta(t-t_0)$ is small.  A more precise condition that the MP model is reasonable if $\eta (\Pi_I + \Pi_R)/r \ll 1$ where $r$ is the early exponential growth rate is also described in the Appendix.

We show convergence of the DVD model to the MP and MFSH heterogeneity model as $\eta \to 0$ or $\eta \to \infty$ in figure~\ref{fig:DVDconv}.  The technical mathematical details showing this convergence are in the Appendix.

\begin{figure}
\includegraphics{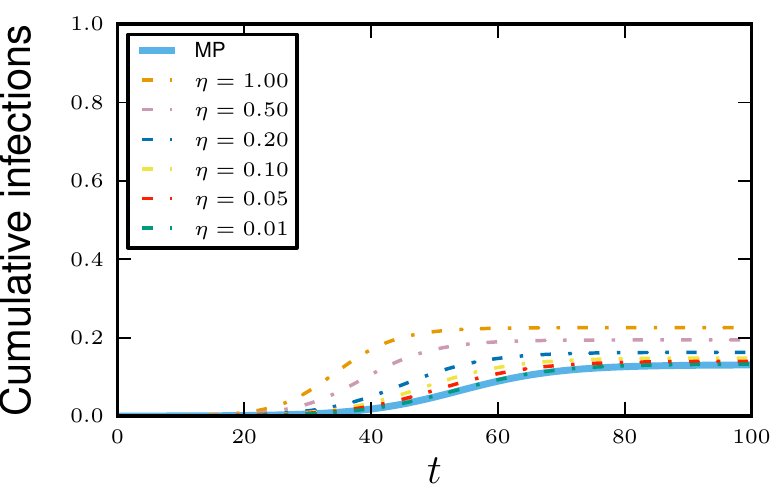}
\includegraphics{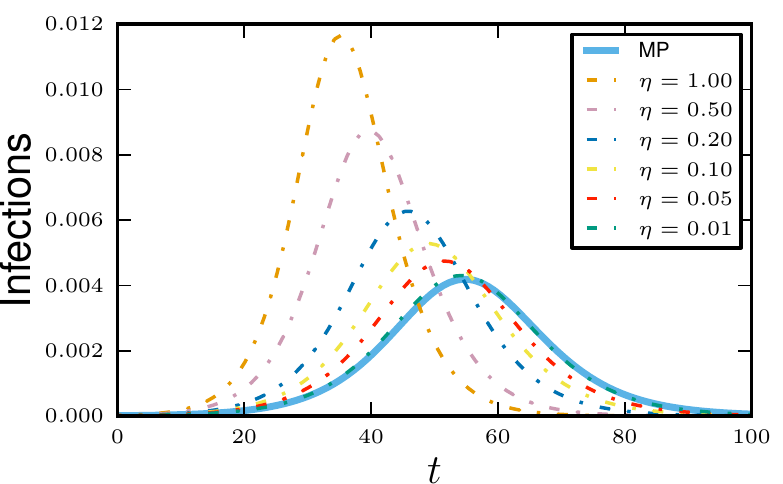}\\
\includegraphics{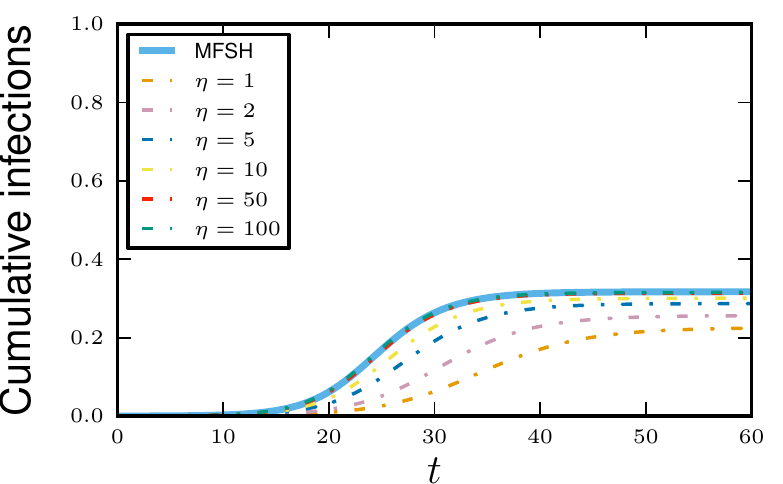}
\includegraphics{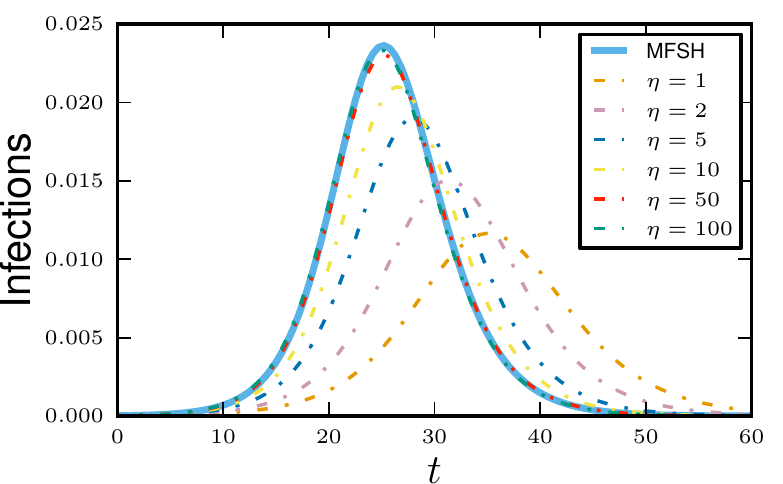}\\
\caption{\textbf{Convergence of DVD to MP and MFSH models}.  We consider the DVD model with varying values of $\eta$.  The probability density function for the expected degree is $\rho(\kappa) = e^\kappa/[\exp(10)-1]$ for $\kappa \in (0,10)$ and $0$ otherwise.  This yields $\Psi(x)=[1-\exp(-10(1-x))]/[10(1-x)]$.  We take $\beta = 0.2$ and $\gamma = 1$.  As $\eta$ decreases (top), the MP model results, while as $\eta$ increases (bottom) the MFSH model results.  Note the difference in axes from top to bottom.}
\label{fig:DVDconv}
\end{figure}

\subsection{Actual degree models}
For our second class of models, we assume that each node has an actual degree $k$ assigned using the probability mass function $P(k)$.  The degree must be a non-negative integer.  Each individual is given $k$ stubs, and at any time those stubs may join in pairs with stubs of other nodes to form edges.  In most models we assume that the stubs are always in pairs (though the partner may change), so the degree is $k$.  In the Dormant Contact model, we allow stubs to be active or dormant, and thus take $k_m$ [distributed according to $P(k_m)$] to be the maximum degree of a node, with $k_a$ and $k_d$ the active and dormant degrees respectively $k_a+k_d=k_m$.  We use $\theta(t)$ to denote the probability that a stub has not transmitted infection from a neighbor to its node by time $t$.  The probability a node with a given $k$ is susceptible is $\theta^k$, and taking a weighted average over all $k$, we find $S(t) = \psi(\theta(t))$ where
\[
\psi(x) = \sum_k P(k) x^k
\]

\subsubsection{Configuration Model}
The Configuration Model (CM) networks are similar to the MP networks.  In a CM network, the exact degree of an individual is assigned.  A node is given $k$ stubs, assigned using the probability mass function $P(k)$.  Once all stubs are assigned to nodes, stubs are randomly paired into edges.  The resulting network is static.  
We define $\theta$ to be the probability that the neighbor along a random stub from $u$ has not transmitted infection to $u$ (so $1-\theta$ is the probability the neighbor has transmitted).  We find that
\begin{align*}
\dot{\theta} &= - \beta \theta + \beta \frac{\psi'(\theta)}{\psi'(1)} + \gamma (1-\theta)\\
\dot{R} &= \gamma I \, , \qquad\qquad S = \psi(\theta) \, , \qquad\qquad I = 1-S-R
\end{align*}
This is similar to the MP model.  In fact, it is possible to show that a MP network is a special case of the CM networks.  In  MP networks, a node $u$ with expected degree $\kappa_u$ has its actual degree chosen from a Poisson distribution of mean $\kappa_u$ (in the limit of a large network).  Thus in an MP network the probability a node has degree $k$ is $P(k) = \int_0^\infty [e^{-\kappa}\kappa^k/k!] \rho(\kappa) \, \mathrm{d}\kappa$.  Thus $\psi(x) = \sum_k P(k) x^k = \int_0^\infty e^{-\kappa} (\sum \kappa^kx^k/k!) \rho(\kappa)\, \mathrm{d}\kappa = \Psi(x)$, and so the MP model emerges as a special case of the CM model.

\subsubsection{Mean Field Social Heterogeneity [actual degree formulation]}
In the actual degree version of Mean Field Social Heterogeneity (MFSH), each individual has $k$ stubs.   Stubs change neighbors quickly so that the neighbor at any given time has no bearing on who the neighbor is later.   The governing equations are
\begin{align*}
\dot{\theta}&= -\beta \theta +\beta \theta \frac{\theta\psi'(\theta)}{\psi'(1)} - \gamma \theta\ln \theta\\
\dot{R} &= \gamma I \, , \qquad\qquad S = \psi(\theta) \, , \qquad\qquad I = 1-S-R
\end{align*}

Using techniques similar to those for the expected degree formulation of the MFSH model, we can show that if all degrees are the same, then this reduces to the MA model.  Similarly, if the degrees are sufficiently close to the mean degree in the sense that $\ave{K^4}/\ave{K}^4 \to 1$ and both $\gamma$ and $\beta \ave{K}$ are fixed, then the solution again converges to that of the MA model.   In section~\ref{sec:mfshequiv} we show that in fact this model is equivalent to the expected degree formulation of the  MFSH model.



\subsubsection{Dynamic Fixed-Degree }
In the Dynamic Fixed-Degree (DFD) model a node is given $k$ stubs, which are paired with stubs of other nodes into edges.  As time progresses, an edge may break, and the freed stubs immediately form edges with stubs from other edges that break at the same time, a process we refer to as ``edge swapping''.  The rate any edge breaks is $\eta$.  The resulting equations are
\begin{align*}
\dot{\theta} &= -\beta \phi_I \, ,\\
\dot{\phi}_S &= - \beta \phi_I \phi_S\frac{\psi''(\theta)}{\psi'(\theta)}  + \eta\theta\pi_S - \eta \phi_S\, ,  \\
\dot{\phi}_I &= \beta \phi_I \phi_S\frac{\psi''(\theta)}{\psi'(\theta)} + \eta \theta \pi_I - (\beta  + \gamma + \eta) \phi_I \, ,  \\
\dot{\pi}_R &= \gamma \pi_I \, , \qquad\qquad \pi_S = \frac{\theta\psi'(\theta)}{\psi'(1)} \, ,\qquad\qquad 
\pi_I = 1-\pi_R-\pi_S\, ,  \\ 
\dot{R} &= \gamma I \, ,\qquad\qquad S(t) = \psi(\theta) \, ,\qquad\qquad I(t) = 1-S-R \, . 
\end{align*}
The new variables $\phi_S$ and $\phi_I$ represent the probability that a stub has not transmitted infection from any neighbor to its node and is currently connected to a susceptible or infected neighbor respectively.  The variables were not needed in the earlier models because we can solve for them explicitly in those cases.  The new variables $\pi_S$, $\pi_I$, and $\pi_R$ represent the proportion of all stubs that belong to susceptible, infected, or recovered nodes.  A newly formed edge connects a node to a susceptible, infected, or recovered node with probabilities $\pi_S$, $\pi_I$, and $\pi_R$ respectively.

The DFD model plays the same role in the actual degree case that DVD model played in the expcted degree case.  It experiences similar limiting behavior.  If $\eta/\beta$ is large, we recover the MFSH model.  Alternately the CM is an accurate approximation so long as $\eta(t-t_0)$ is small where $t_0$ is a time around when the epidemic begins to infect significant numbers.  Again, a more precise condition that $\eta (\pi_I+\pi_R)/r \ll 1$ where $r$ is the early exponential growth rate is described in the Appendix.

Figure~\ref{fig:DFDconv} shows the convergence of this model to the CM and MFSH models as $\eta \to 0$ or $\eta \to \infty$.

\begin{figure}
\includegraphics{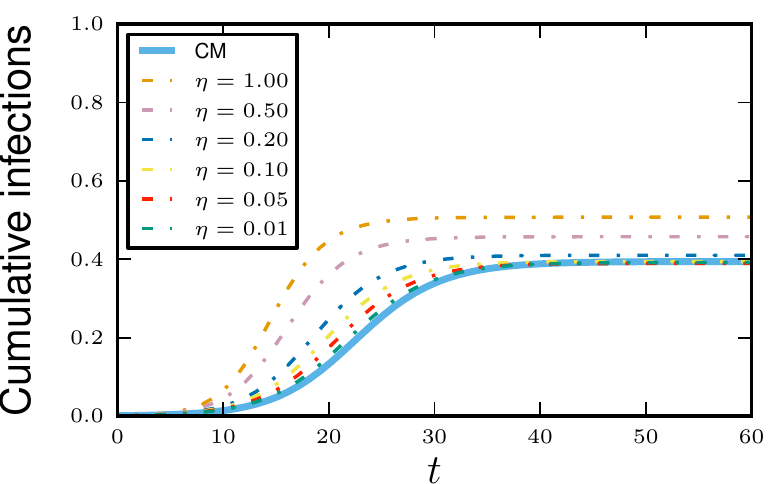}
\includegraphics{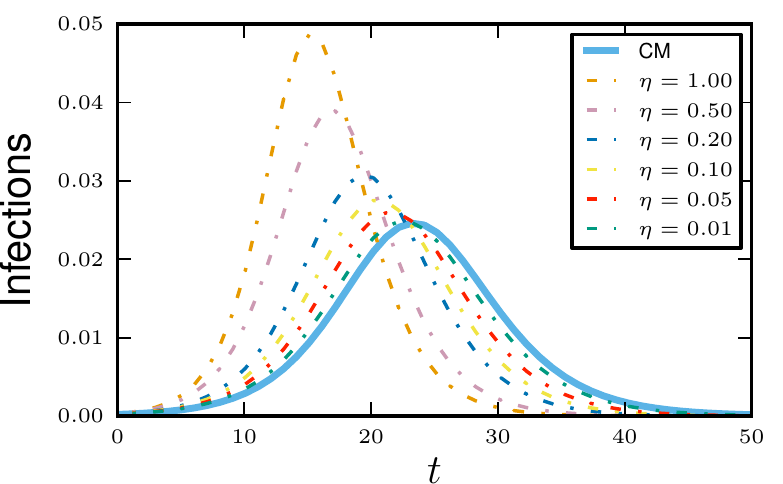}\\
\includegraphics{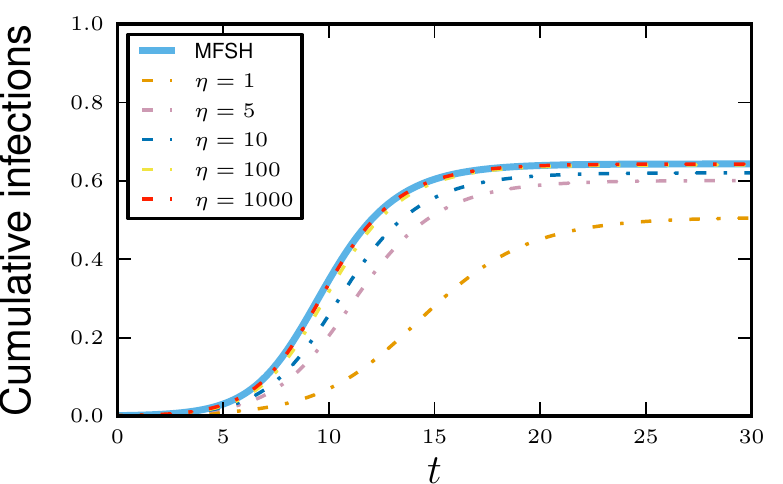}
\includegraphics{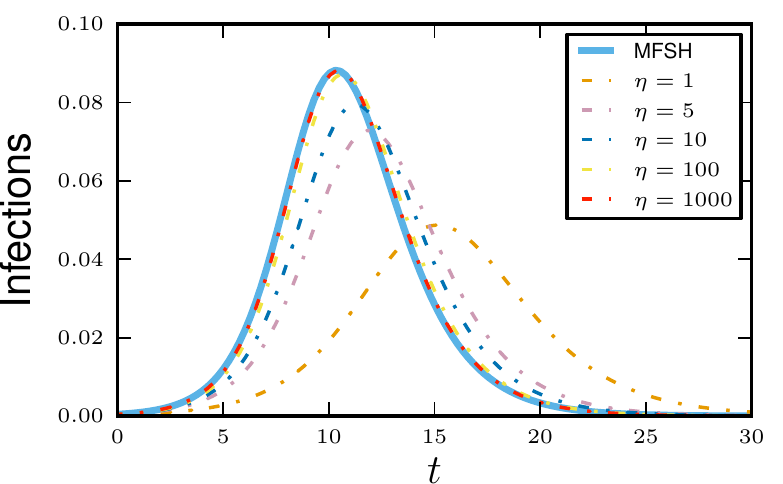}
\caption{\textbf{Convergence of DFD to CM and MFSH models}.  We consider the DFD model with varying values of $\eta$.  The degrees are $k=6$, \ $k=8$, and $k=10$, with probability $1/3$ each.  This yields $\psi(x) = (x^6+x^8+x^{10})/3$.   We take $\beta = 0.2$ and $\gamma = 1$.  As $\eta$ decreases (top), the CM model results, while as $\eta$ increases (bottom) the MFSH model results.  Note the difference in axes from top to bottom.}
\label{fig:DFDconv}
\end{figure}

\subsubsection{Dormant Contacts}

\begin{figure}
\includegraphics{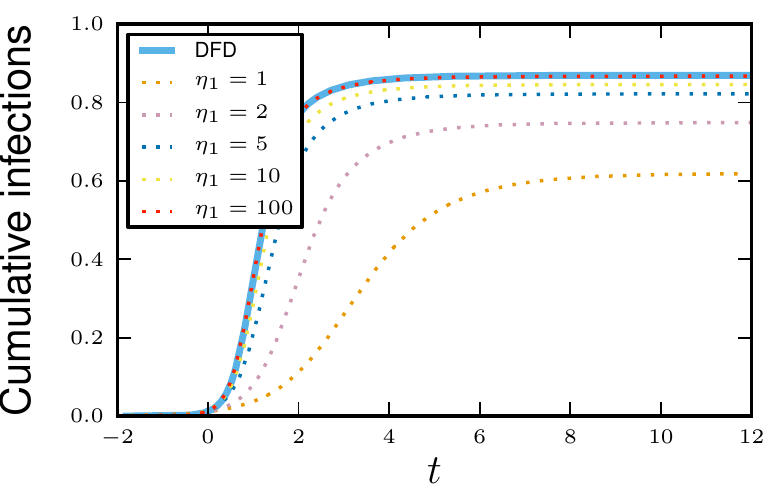}
\includegraphics{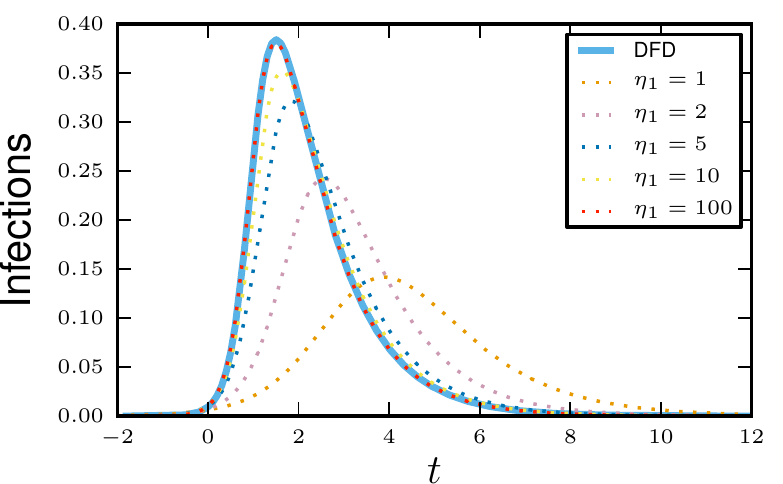}\\
\includegraphics{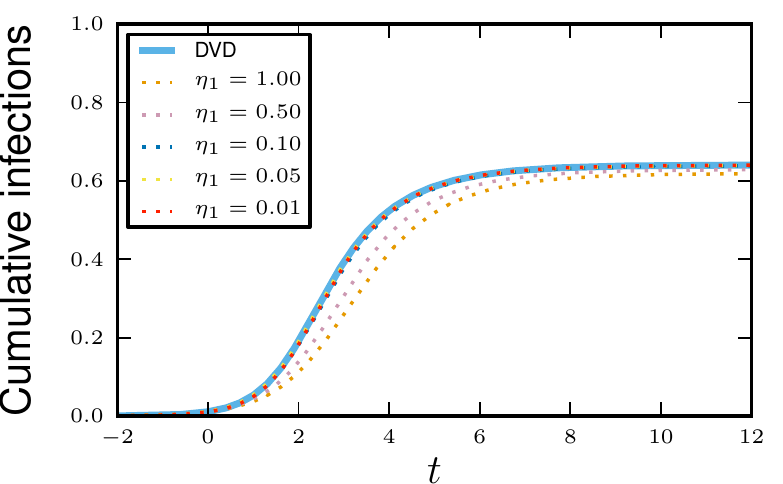}
\includegraphics{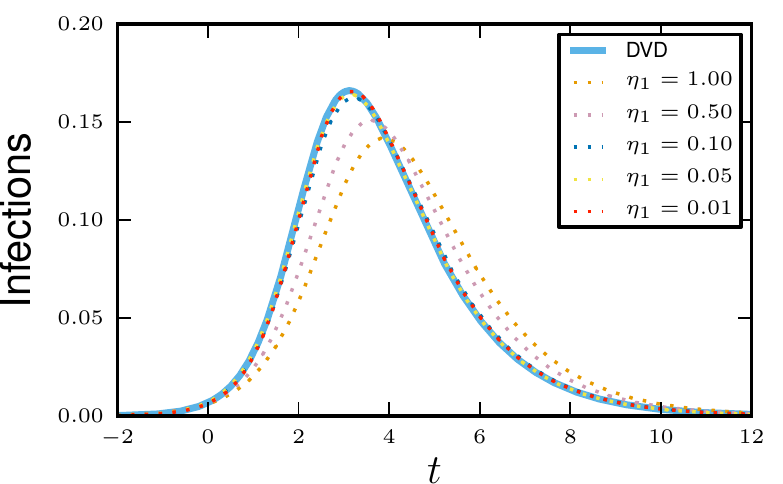}\\
\includegraphics{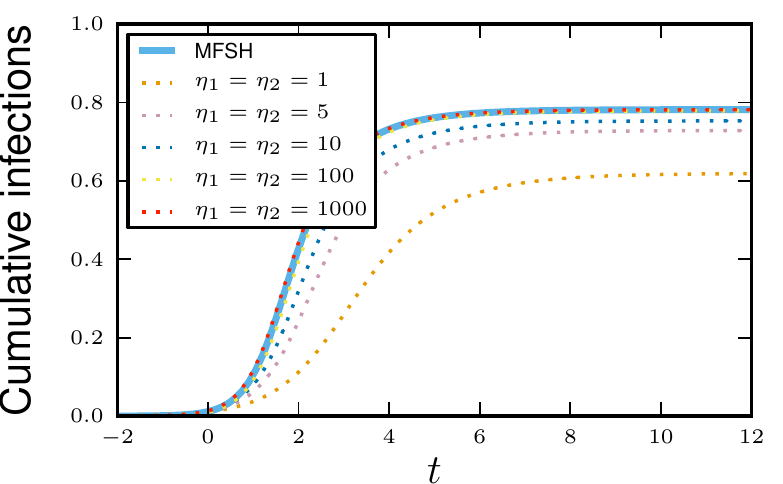}
\includegraphics{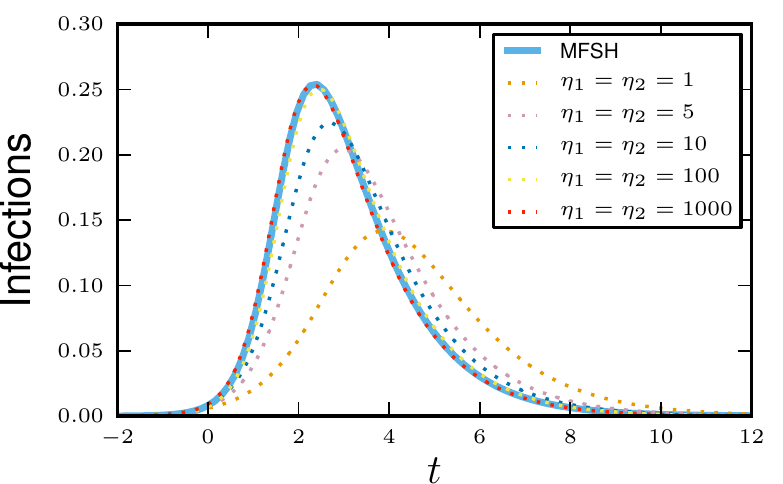}
\caption{\textbf{Convergence of DC to DFD, DVD, and MFSH models}.  We take $\beta=\gamma=1$.  (top) We consider the convergence of the DC model to the DFD model as $\eta_1$ increases.  We take $\psi(x) = (x^2+x^8)/2$.  For the DC model we take $\eta_2=1$ and vary $\eta_1$.   As $\eta_1$ increases, the DC model converges to the DFD model result for $\eta=1$.  (middle) We consider the convergence of the DC model to the DVD model as $\eta_1$ decreases and the degrees increase to compensate.  For the DC model, we have $\psi(x) = (x^{1/\xi} + x^{4/\xi})/2$ where $\xi = \eta_1/(\eta_1+\eta_2)$.  We use $\eta_2=1$ and vary $\eta_1$.  For the DVD model, we take $\eta=1$ and $\Psi(x) = (\exp[-(1-x)] + \exp[-4(1-x)])/2$.  (bottom)  We consider the convergence of the DC model to the MFSH model with $\eta_1=\eta_2$ and $\eta_1,\eta_2 \to \infty$.  We take $\psi(x) = (x^2+x^8)/2$.  The MFSH model has the same $\psi$, but uses $\beta\xi=\beta/2$ as the transmission rate.  Note that for all three calculations, the $\eta_1=1$ curves are the same.}
\label{fig:DClimits}
\end{figure}

We finally move to the Dormant Contact (DC) model which captures all the previous expected and actual degree models as limiting cases.  In the DC model, each node is given $k_m$ stubs [with $k_m$ chosen from $P(k_m)$].  However, only a fraction of them are active.   At any given time, the node will have $k_a$ active stubs and $k_d$ dormant stubs, so $k_m=k_a+k_d$ is the maximum number of active stubs.  Active stubs become dormant at rate $\eta_2$ and dormant stubs become active at rate $\eta_1$.  We take $\psi(x) = \sum_{k_m} P(k_m) x^{k_m}$.  The governing equations are
\begin{align*}
\dot{\theta} &= -\beta \phi_I\\
\dot{\phi}_S &= - \beta \phi_I \phi_S \frac{\psi''(\theta)}{\psi'(\theta)} + \eta_1 \frac{\pi_S}{\pi}\phi_D - \eta_2\phi_S\\
\dot{\phi}_I &= \beta \phi_I\phi_S\frac{\psi''(\theta)}{\psi'(\theta)} + \eta_1 \frac{\pi_I}{\pi} \phi_D - (\eta_2+\beta+\gamma)\phi_I \\
\dot{\phi}_D &= \eta_2(\theta-\phi_D) - \eta_1 \phi_D\\
\dot{\xi}_R &= -\eta_2 \xi_R + \eta_1 \pi_R +\gamma \xi_I \, , \qquad \xi_S = (\theta-\phi_D)\frac{\psi'(\theta)}{\psi'(1)} \, , \qquad \xi_I = \xi- \xi_S - \xi_R\\
\dot{\pi}_R &= \eta_2 \xi_R-\eta_1\pi_R + \gamma \pi_I\, , \qquad \pi_S = \phi_D \frac{\psi'(\theta)}{\psi'(1)}  \, , \qquad \pi_I = \pi - \pi_S - \pi_R\\
\xi &= \frac{\eta_1}{\eta_1+\eta_2} \, , \qquad \pi = \frac{\eta_2}{\eta_1+\eta_2}\\
\dot{R} &= \gamma I \, , \qquad\qquad S = \psi(\theta) \, , \qquad\qquad I = 1-S-R 
\end{align*}
The new variable $\phi_D$ represents the probability that a stub has not transmitted infection to its node and is currently dormant.  The variables $\pi_S$, $\pi_I$, and $\pi_R$ now measure the proportion of all stubs which are both dormant and belong to a susceptible, infected, or recovered node, with $\pi=\pi_S+\pi_I+\pi_R=\eta_2/(\eta_1+\eta_2)$ the probability a stub is dormant.  The ratios $\pi_S/\pi$, $\pi_I/\pi$, and $\pi_R/\pi$ give the probabilities that a newly formed edge connects to a susceptbile, infected, or recovered node.  The variables $\xi_S$, $\xi_I$, and $\xi_R$ give the probabilities that a stub is active and belongs to a susceptible, infected, or recovered node, with $\xi = \xi_S+\xi_I+\xi_R=\eta_1/(\eta_1+\eta_2)$ the probability a stub is active.

It is relatively straightforward to see that if $\eta_1$ is much larger than $\eta_2$, then the proportion of time a stub is dormant is tiny, $\pi \ll 1$.  Consequently at any moment a node is expected to have $k_a=(1-\pi)k_m \approx k_m$ active contacts.  As $\eta_2/\eta_1$ shrinks, this approximation improves.  So in this limit, the DC model should reduce to the DFD model with edges breaking at rate $\eta_2$.  Indeed, in the equations above, if we take $\eta=\eta_2$ and assume $\eta_1 \gg \eta$, then $\phi_D$ is negligibly small, so the $\phi_D$ terms drop out of the $\phi_S$ and $\phi_I$ equations.  The values of $\pi_S$ and $\pi_I$ both go to zero as $\eta_1$ grows, but a little more care shows that $\pi_S/\pi$ approaches $\xi_S$ and $\pi_I/\pi$ approaches $\xi_I$ and the $\xi$ variables solve the same equations in this limit as the $\pi$ variables in the DFD equations.   Thus the DC equations reduce to the DFD equations in this limit.  More details are in the Appendix.

In the opposite limit, if $\eta_2$ is large compared to $\eta_1$, then stubs spend most time dormant, $\xi \ll 1$.  If the number of stubs is sufficiently large, the expected number of active stubs $\kappa=\E[k_a]=\xi k_m$ will not be negligible.  The variation in the number of active stubs will be significant relative to its expected value $\kappa$.  However, $\kappa \ll k_m$ so the number of dormant stubs $k_d = k_m-k_a$ can be approximated as $k_m\pi$.   The rate that dormant stubs become active is $\eta_1$, so the rate at which the node forms new edges is approximately $\eta_1 k_d= \eta_1 k_m\pi$ which itself is $\eta_2 \kappa$, while each existing edge breaks at rate $\eta_2$.  This is the assumption underlying the DVD model.  A careful analysis of the equations show that indeed we can reduce them to the DVD model if $\eta_2 \gg \eta_1$.  More details are in the Appendix.

Alternately, we can have the DC model converge to the MFSH model if $\ave{K} \to \infty$, as long as $\beta\ave{K}$ remains fixed.  In this case we find that the MFSH model is a good approximation, using $\hat{\beta} = \beta \xi$ as the transmission rate.  The underlying argument of this is that as $\ave{K}$ increases, the probability of transmission per edge decreases.  Consequently, it becomes unimportant whether the edge has long or short duration because it is incredibly rare for the disease to try to transmit along the same edge twice.  The total infectiousness of an individual is then given by the number of active edges times the per-edge infection rate.  We can assume that only a proportion $\xi$ of the stubs are active at any time, and we can absorb this into $\beta$.  So an individual with $k$ stubs causes transmission at rate $\hat{\beta} k$ where $\hat{\beta} = \xi \beta$.  This is independent of how (or if) the edges are changing in time.

Figure~\ref{fig:DClimits} shows how the DC model reduces in the various limits.


\subsection{Further Analysis of Mean Field Social Heterogeneity models}
In this section we show that the two MFSH models are equivalent.  We also show that all the non-Mass Action models described here reduce to the MFSH model if the average number of contacts is large, and the probability of infection per edge scales like $1/\ave{K}$.  Technically, our results show that there is no difference between the results of any of the (non-MA) models in this limit, so any of these models would be appropriate, but generally the equations of the MFSH model are simplest so we use it.

\subsubsection{Equivalence of Mean Field Social Heterogeneity formulations}
\label{sec:mfshequiv}

There are two formulations of the MFSH model, one with expected degrees and the other with actual degrees.  

In the expected degree formulation, each individual has an expected degree $\kappa$.  On average, at any given moment of time the individual has $\kappa$ contacts.  The exact number of contacts may vary from moment to moment, but how many contacts exist at one moment and who those contacts are with are independent from any other moment.  Thus over a short period of time, we may safely assume that the total number of contacts with infected individuals matches the expected number.

In the actual degree formulation, each individual is assigned an actual degree $k$ and given $k$ stubs.  At any time each of those $k$ stubs forms an edge with a stub from another node, but who the stub connects to changes rapidly.  Again, over a short period of time, we may assume that the number of contacts with infected individuals matches the expected number.

\paragraph{Actual degree formulation as a special case of expected degree formulation} 
Because contacts are very short, over any time interval, the total contact time is well-approximated based on the expected amount of contact at any given moment. Thus whether the individuals have the same number of contacts at each moment, or whether the amount varies from moment to moment, the effect over any macroscopic time interval is the same. Thus the actual degree model should be a special case of the expected degree model.  In the expected degree formulation, the probability that a node with $\kappa$ expected contacts is susceptible is $\exp[-\kappa(1-\Theta)]$ and in the actual degree formulation the probability that a node with $k$ contacts is susceptible is $\theta^k$.  If $\kappa=k$, we expect these to equal, and so we anticipate $\theta = \exp(\Theta-1)$, or $\Theta = 1+\ln \theta$.  Plugging this into the expected degree equations, we arrive at the actual degree equations.  So the two methods are equivalent subject to a change in variables.


\paragraph{Expected degree formulation as a limiting case of the actual degree formulation}

The fact that the actual degree formulation leads to the expected degree formulation is based on similar reasoning.  The underlying additional idea is that any continuous distribution can be approximated by a sufficiently well-refined discrete distribution.  We then need a way to take a well-refined discrete distribution and rescale it so that all the probability is massed at integer values.  To do this we make the observation that we can multiply everyone's (expected or actual) degree by $L$ with no impact on the epidemic so long as we also divide the transmission rate by $L$.  

To make this more precise, consider a given $\beta$ and $\rho(\kappa)$ and assume $L$ is large.  We approximate the continuous distribution of $\kappa$ using a discrete distribution where the probability of $k/L$ is $\int_{k/L}^{(k+1)/L} \rho(\kappa) \, \mathrm{d}\kappa$.  Increasing $L$ gives a finer scale approximation.  We then multiply every $k/L$ by $L$, to get $P(k) = \int_{k/L}^{(k+1)/L} \rho(\kappa) \, \mathrm{d}\kappa$ where $k$ is an integer.  So long as we divide $\beta$ by $L$, the spread of the epidemic on a population with the given $P(k)$ and the rescaled $\beta$ will closely approximate the epidemic spread in the original expected degree population, with the approximation improving as $L\to\infty$.  We find that the actual degree equations converge to the expected degree equations as $L \to \infty$.  Full details are in the Appendix.

\begin{figure}
\includegraphics{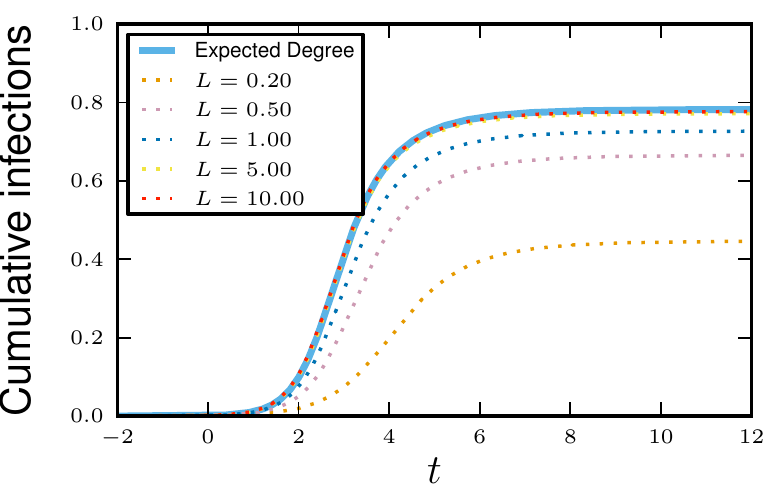}
\includegraphics{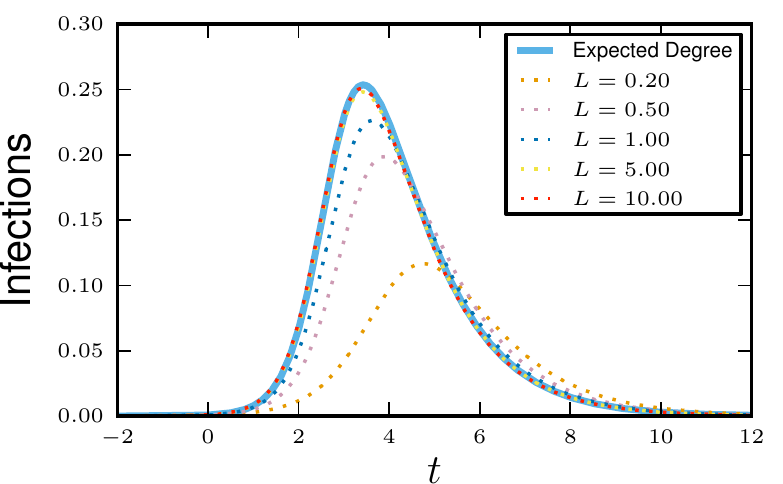}
\caption{\textbf{Convergence of actual degree formulation of MFSH to expected degree formulation}.  The actual degree formulation of the MFSH model for the appropriate discrete distribution (described in text) converges to the expected degree formulation of the MFSH model with contact rates chosen uniformly from $0$ to $10$.}
\label{fig:MFSH_equiv}
\end{figure}

As an example we take the uniform distribution from $0$ to $10$ for $\kappa$.  So $\Psi(x) = (1-\exp[-10(1-\Theta)])/[2(1-\Theta)]$.  We initially start with a discrete distribution on the integers $0$ through $9$ with weight $1/10$ on each.  So our initial $\psi(x)$ is $\sum_{k=0}^9 x^k/10 = (x^{10}-1)/10(x-1)$ with $\beta = \hat{\beta}$.  We then refine the distribution.  For given $L$ we have $\psi(x)=\sum_{k=0}^{10L-1} x^k/10L = (x^{10L}-1)/[10L(x-1)]$, and we take $\beta = \hat{\beta}/L$.  Convergence is shown in figure~\ref{fig:MFSH_equiv}.

\subsubsection{Reduction of all models to Mean Field Social Heterogeneity model at large average degree}
One particularly important limit corresponds to people having many contacts, but a low probability of transmitting per contact before recovering.  That is, $\ave{K}$ is large, and $\beta/\gamma$ is comparably small.  In this limit, we expect that the duration of contact becomes unimportant, because the infection is unlikely to cross an edge more than once so it has no way to know how long the edge lasts.  Heterogeneity in contact levels may still play an important role.  If the heterogeneity is sufficiently small, it is reasonable to expect that the MA SIR model is appropriate.

In section~\ref{sec:DC2MFSHrigor} we rigorously derive the MFSH model from the DC model assuming $\ave{K} \to \infty$ with  fixed $\beta\ave{K}$ and fixed $\gamma$.  A similar proof will apply for the other models, or we can simply argue that the DC model reduces to all of the others, and careful attention to detail shows that they inherit this limiting case.

\begin{figure}
\includegraphics{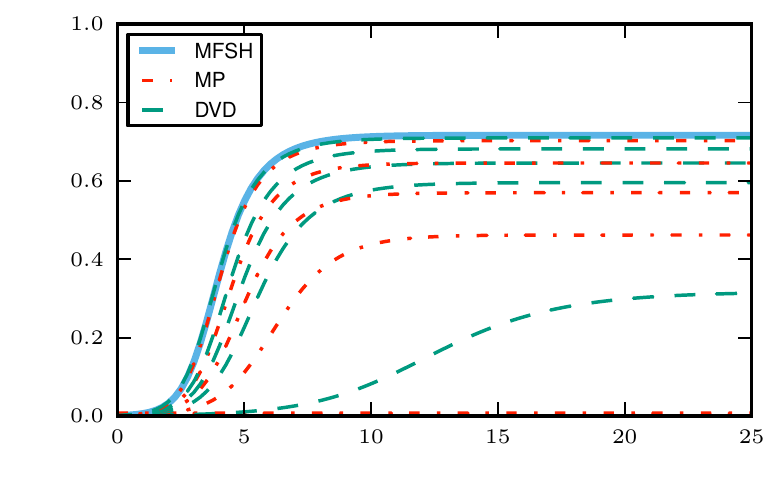}
\includegraphics{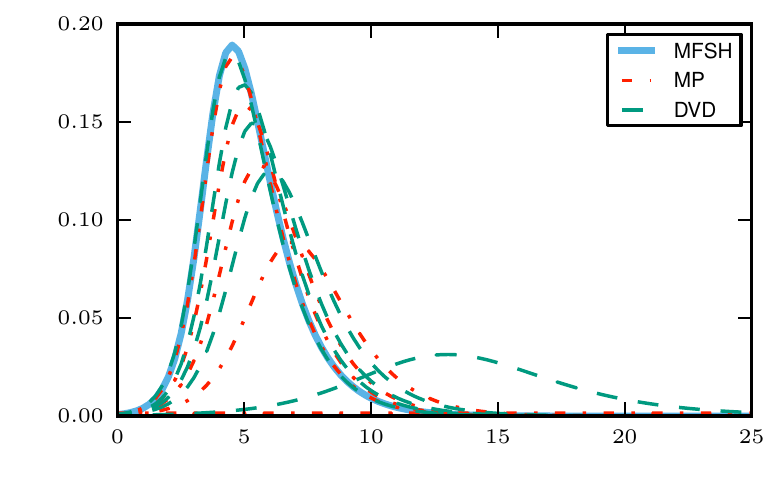}\\
\includegraphics{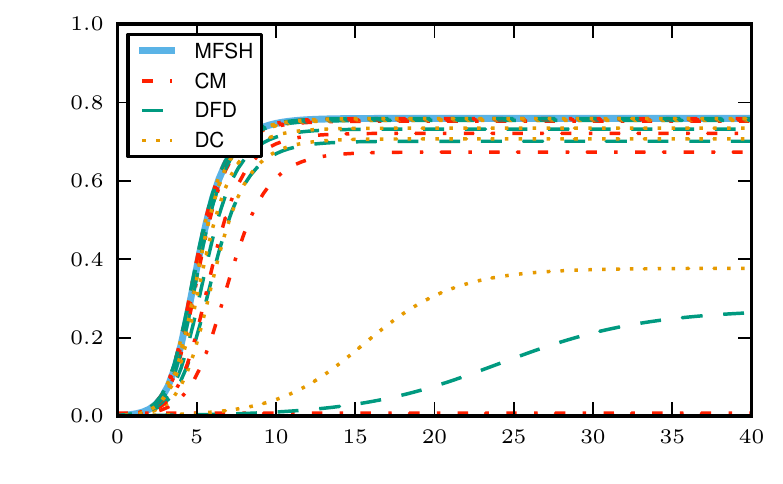}
\includegraphics{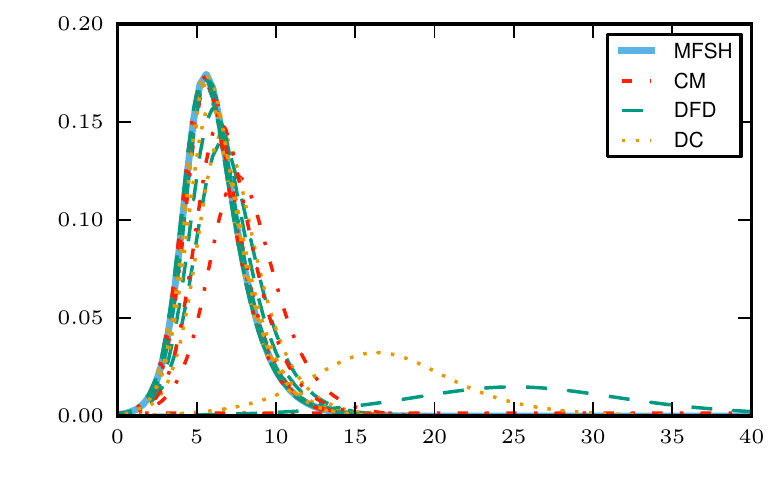}
\caption{\textbf{Convergence to MFSH}. (top) Convergence of the Expected Degree models assuming a uniform distribution from $0$ to $2\ave{K}$.  (bottom) Convergence of the Actual Degree models assuming a third each with $\ave{K}/2$, $\ave{K}$, and $3\ave{K}/2$ stubs.  The solid curves correspond to $\ave{K}=1$ (top) and $\ave{K}=2$ (bottom).  The dotted curves correspond to various larger values of $\ave{K}$.  Because $\beta$ decreases to balance the increase in $\ave{K}$, the MFSH curves are the same for all $\ave{K}$.}
\end{figure}

\subsubsection{Reduction of MFSH to MA model if $\ave{K^4}/\ave{K}^4\to 1$ and $\beta\ave{K}$ fixed}
\label{sec:MFSH2SIR}
It is straightforward to show that the MFSH model becomes SIR model if every node has the same contact rate.  However more generally, we would expect that if the contact rates are sufficiently close together, the model should behave like the SIR model.  In fact this holds if $\ave{K^4}/\ave{K}^4 \to 1$ with $\beta\ave{K}$ fixed.  Intuitively, if the number of nodes with higher or lower contact rates is very small, their contribution to the spread of the disease is not significantly different from the contribution of an average node, and so the disease should spread as if the contact rate were homogeneous. We do not have a good intuitive explanation for why the precise condition relies on $\ave{K^4}/\ave{K}^4$.  However, we prove it rigorously in section~\ref{sec:MFSH2MArigor}.

We expect this case to be particularly relevant if the average degree is large and infectiousness is low.  Combining this with the previous result, we conclude that all models converge to the MA model under appropriate conditions as $\ave{K} \to \infty$.


\section{Rigorous proof of convergence to Mass Action equations}
\label{sec:rigorous}

In this section, we give a rigorous proof of one of the most significant results.  Namely, if $\ave{K}$ is large, but $\ave{K^4}/\ave{K^4}$ is approximately $1$ and $\beta\ave{K}$ and $\gamma$ are small compared to $\ave{K}$, then regardless of which model we use, the result is well-approximated by the MA equations.  

We first show some examples, and then provide the details needed to prove the result for the DC model.  The simpler models can be derived in the same method (though in some cases the proof will be simpler).  
For our proof, we assume we have a sequence of populations and diseases indexed by $n$ such that $\ave{K} \to \infty$, \ $\ave{K^4}/\ave{K}^4 \to 1$ with $\xi\beta\ave{K}$ fixed. We show that this converges to the mass action equations using $\hat{\beta}=\xi \beta\ave{K}$.

We show this in two steps.First, we show that in the limit of large $\ave{K}$ and constant $\beta\xi\ave{K}$, the DC equations converge to the MFSH equations. We then show that the MFSH equations converge to the MA equations if additionally $\ave{K^4}/\ave{K}^4 \to 1$.

\subsection{Example}
\begin{figure}
\includegraphics[width=0.48\textwidth]{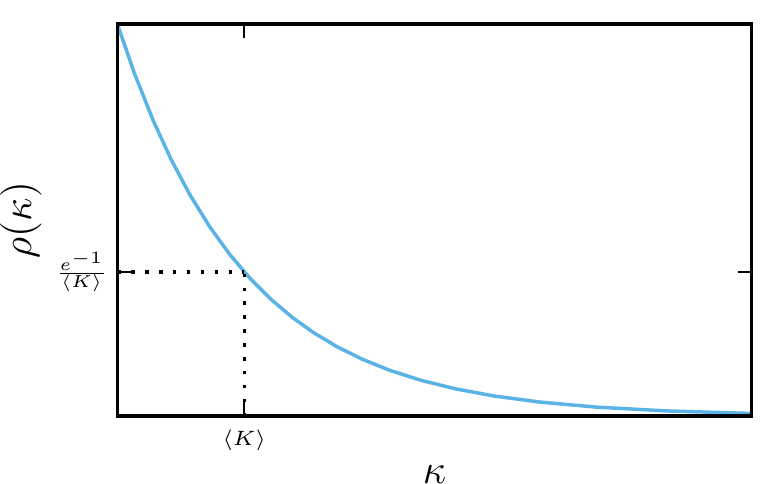}
\includegraphics[width=0.48\textwidth]{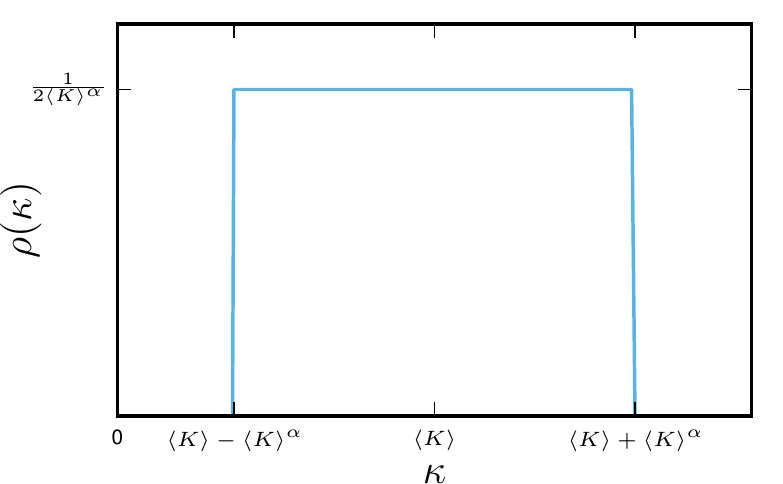}
\caption{The expected degree distributions for our examples of convergence as $\ave{K} \to \infty$, with and without $\ave{K^4}/\ave{K}^4 \to 1$.}
\label{fig:deg_dist}
\end{figure}

\begin{figure}
\includegraphics[width=0.48\textwidth]{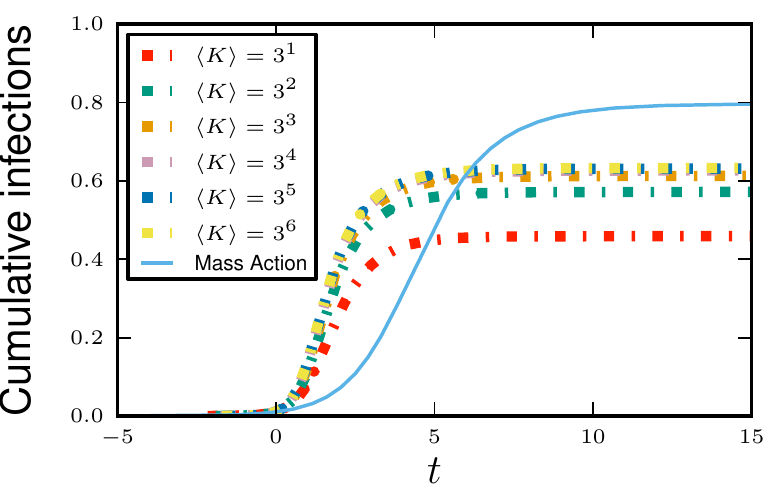}
\includegraphics[width=0.48\textwidth]{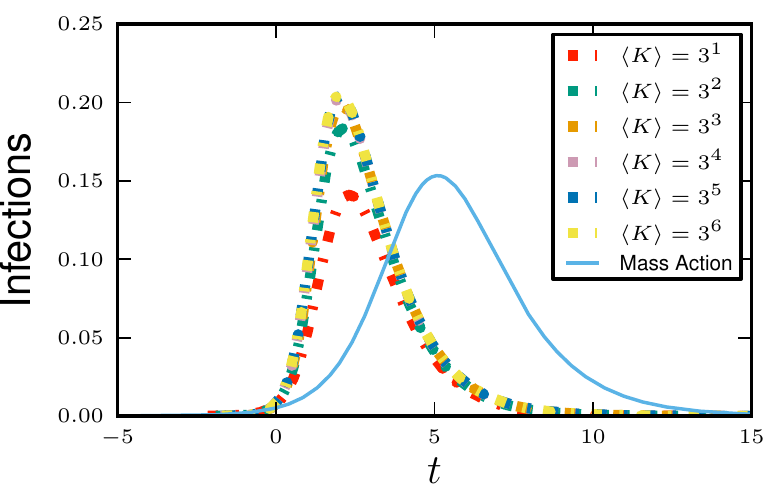}\\
\caption{\textbf{Sample convergence as $\ave{K} \to \infty$.}  We use the exponential distribution described in figure~\ref{fig:deg_dist}.  The distribution of expected degrees decays exponentially and $\ave{K^4}/\ave{K}^4$ does not approach $1$ as $\ave{K}$ grows.  The solutions appear to converge, but they do not converge to the MA SIR model.}
\label{fig:expconv2SIR}
\end{figure}
As examples, we consider several different population structures using the MP model in figure~\ref{fig:deg_dist}.  In the first, we take an exponential distribution of expected degrees.  To vary the average degree, we change the decay rate of the exponential. 
\[
\rho(\kappa) = \frac{e^{-\kappa/\ave{K}}}{\ave{K}}
\]
This does not satisfy the conditions that $\ave{K^4}/\ave{K}^4 \to 1$.  In the other examples, we take a uniform distribution, with expected degrees chosen uniformly from $(\ave{K}-\ave{K}^\alpha,\ave{K}+\ave{K}^\alpha)$.  That is,
\[
\rho(\kappa) = \begin{cases} 0 & \kappa<\ave{K}-\ave{K}^\alpha\\
 \frac{1}{2\ave{K}^\alpha} & \ave{K}-\ave{K}^\alpha<\kappa<\ave{K}+\ave{K}^\alpha\\
0 & \kappa > \ave{K} + \sqrt{\ave{K}}
\end{cases}
\]
This satisfies the condition for any $\alpha < 1$.  Using $\ave{\hat{K}^4}$ to denote the average of the $4$th power of the expected degree\footnote{Our condition that $\ave{K^4}/\ave{K}^4 \to 1$ applies equally for the fourth power of actual or expected degree since at leading order they are the same.}, we have $\ave{\hat{K}^4} = \ave{K}^4 + 2\ave{K}^{2+2\alpha} + \ave{K}^{4\alpha}/5$.  However, the condition fails for $\alpha=1$.  The distribution would not make sense for $\alpha>1$ since some nodes would have negative expected degree. 

\begin{figure}
\begin{center}
\includegraphics[width=0.48\textwidth]{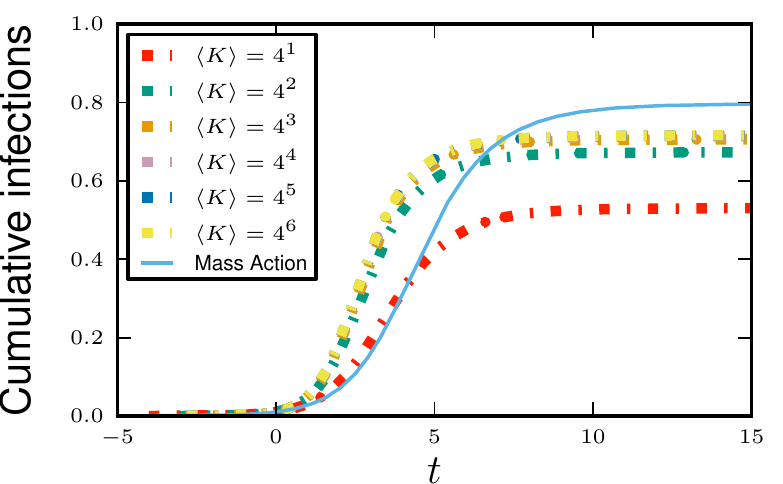}\hfill
\includegraphics[width=0.48\textwidth]{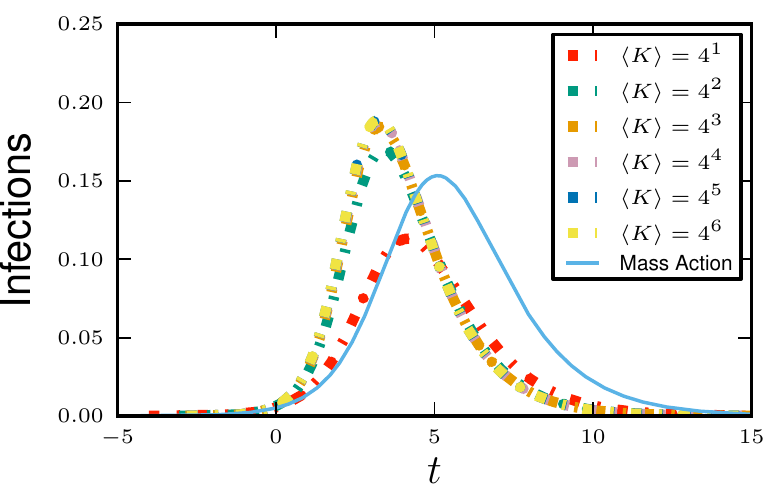}\\[-5pt]
$\alpha = 1$\\[10pt]
\includegraphics[width=0.48\textwidth]{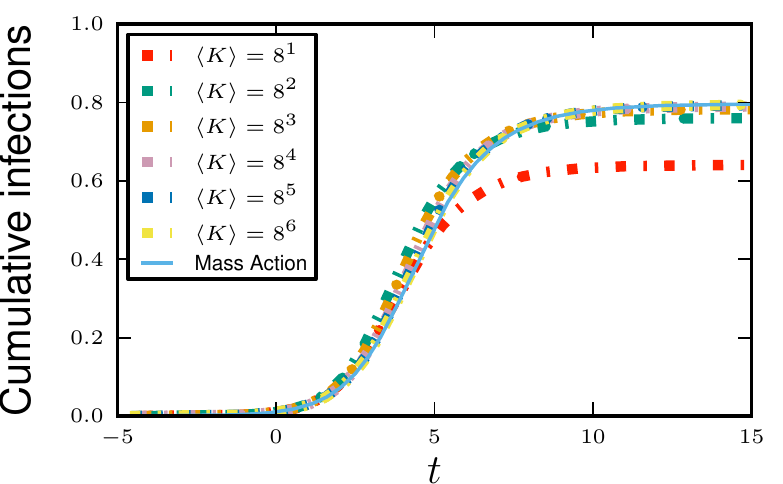}\hfill
\includegraphics[width=0.48\textwidth]{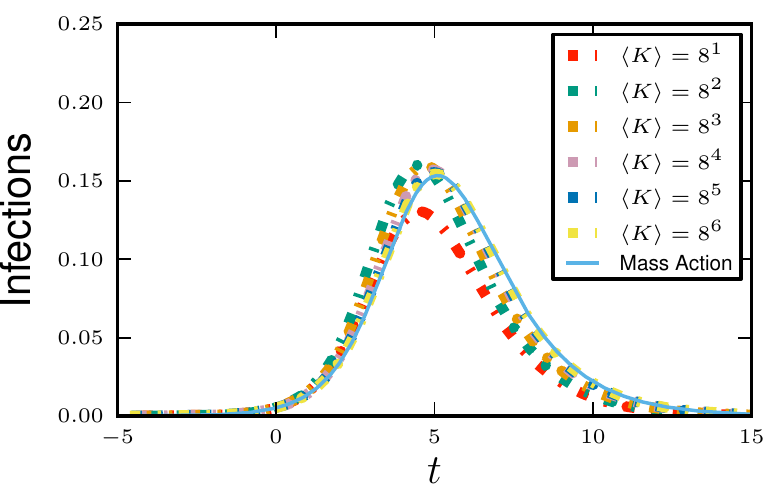}\\[-5pt]
$\alpha = 0.85$\\[10pt]
\includegraphics[width=0.48\textwidth]{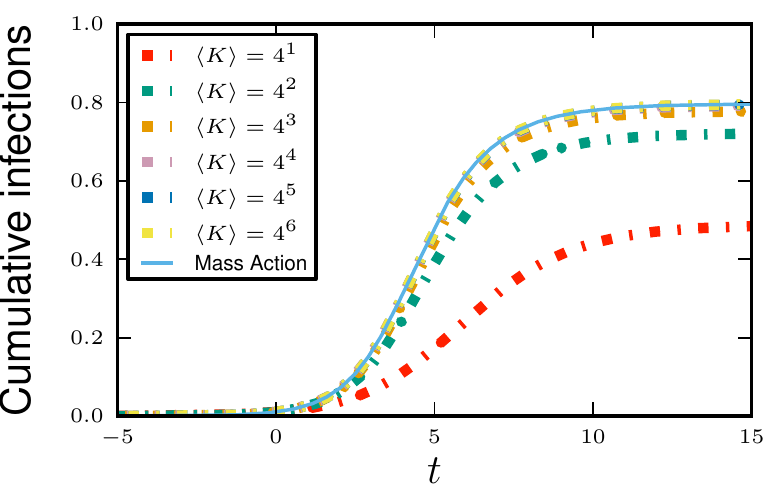}\hfill
\includegraphics[width=0.48\textwidth]{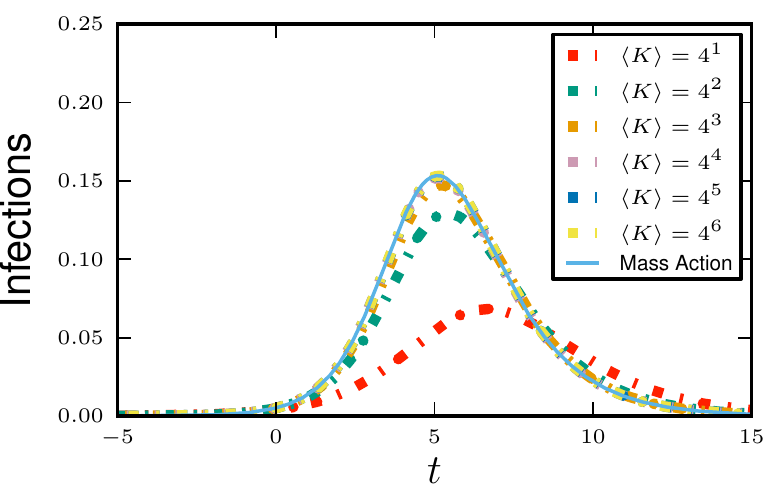}\\[-5pt]
$\alpha = 0.5$
\end{center}
\caption{\textbf{Sample convergence as $\ave{K} \to \infty$}.  We use the uniform distribution described in figure~\ref{fig:deg_dist}.  In the first, second, and third rows, we use $\alpha = 1, 0.85, 0.5$ respectively.  The condition fails for $\alpha = 1$ and the solutions do not converge to the MA model, but they converge to the MA model for smaller $\alpha$.}
\label{fig:conv2SIR}
\end{figure}

For all distributions, we take $\beta = 2/\ave{K}$ and $\gamma = 1$.  The corresponding MA model has $\hat{\beta} = 2$.
In figure~\ref{fig:expconv2SIR} we plot the results for the exponential distribution.  This distribution does not satisfy the conditions, and we see that the solutions do converge, but not to the MA model.  In figure~\ref{fig:conv2SIR}, we take the uniform distribution from $\ave{K}-\ave{K}^\alpha$ to $\ave{K}+\ave{K}^\alpha$.   We again see that when the conditions are not satisfied, the solutions may still converge, but not to the corresponding MA model.  However, when the conditions are satisfied, the solutions converge to the MA model.

\subsection{DC to MFSH}
\label{sec:DC2MFSHrigor}

We will use the following result several times in the proof:  Assume $g(t)$ is a function of time and 
\[
\dot{g} = - c(t) g + f(t)
\]
If $c(t)\geq C>0$ for all time and $|f| \leq F$, and further $|g(0)|\leq F/C$, then for all time $|g(t)| \leq F/C$. To see this, assume that $f$ and $g$ are positive. If $g$ is increasing beyond $F/C$, then $-cg+f$ must be negative, violating the assumption that $g$ is increasing. Similar results hold for negative values of $f$ and $g$.
Our proofs skip some of the intermediate steps. Complete details of all the algebra is in the Appendix.

We want to prove that if degrees increase, but simultaneously the infection rate decreases with $\gamma$ constant then we must arrive at the Mass Action equations.  We do this for the Dormant Contact model.  We introduce the notation $o(1)$ to denote a function whose value goes to zero as $\ave{K} \to \infty$.  We will show that neglecting small terms in the large degree limit the DC model becomes
\begin{align*}
\dot{\theta} &= -\tilde{\beta} \theta \zeta\\
\dot{R} &= \gamma I\,, \qquad S=\psi(\theta) \, , \qquad I = 1-S-R
\end{align*}
where $\zeta = 1 - \theta \psi'(\theta)/\psi'(1) + (\gamma/\tilde{\beta})\ln \theta$. 

We define $\zeta = \pi_I + \xi_I$  This is the proportion of all stubs which belong to infected nodes.  We prove a series of results:
\begin{enumerate}
\item $\theta = 1 + o(1)$.
\item $\phi_D = \pi + o(1)$.
\item $\phi_S = \xi \frac{\psi'(\theta)}{\psi'(1)} + o(1)$.
 \item $\pi_I = \pi \zeta + o(1)$ \qquad $\xi_I = \xi \zeta + o(1)$.
 \item $\phi_I = \xi \theta \zeta + o(1)$
 \item $\dot{\theta} = -\tilde{\beta}\theta \zeta  + \beta o(1)$.
\item $\zeta = 1 - \theta\frac{\psi'(\theta)}{\psi'(1)} + (\gamma/\tilde{\beta})\ln \theta + t o(1)$
 \end{enumerate}
If we drop the error terms in the final two results, then we have the equations governing the Mean Field Social Heterogeneity model in the actual degree case, with $\tilde{\beta}=\beta \xi$ playing the role of the transmission rate.  The existence of the error terms means that at earlier time, the approximation is better, and it can deviate more as time increases.  The solution converges uniformly for any $t$ less than any arbitrary chosen value $T$ as $\ave{K}$ grows.  So for large enough $\ave{K}$, the region over which the MFSH model gives an accurate approximation will include the entire period of the epidemic. 

\begin{enumerate}
\item $\theta = 1 + o(1)$.

We trivially have $1 \geq \theta \geq 1- \beta/(\beta+\gamma)$.  Because $\beta \ave{K}$ is constant, we find $\beta = o(1)$.  Thus $1 \geq \theta \geq 1 + o(1)$ (with a negative error term), and so we can conclude that $\theta = 1 + o(1)$.

\item $\phi_D = \pi + o(1)$.
We show $\phi_D-\pi = o(1)$.
\begin{align*}
\diff{}{t} (\phi_D - \pi) &= \eta_2 (\theta - \phi_D) - \eta_1 \phi_D\\
   &= \eta_2[1 + o(1)] - (\eta_1 + \eta_2) \phi_D\\
  &=  (\eta_1+\eta_2)\pi - (\eta_1+\eta_2)\phi_D + o(1)\\
  &= -(\eta_1+\eta_2)(\phi_D-\pi) + o(1)
\end{align*}
In the next to last step we used the fact that $\pi = \eta_2/(\eta_1+\eta_2)$ so $\eta_2 = (\eta_1+\eta_2)\pi$.  So $\phi_D - \pi$ is at most $o(1)/(\eta_1+\eta_2) = o(1)$.

 \item $\phi_S = \xi \frac{\psi'(\theta)}{\psi'(1)} + o(1)$.

We have
\begin{align*}
\diff{}{t} \left[\phi_S - \xi \frac{\psi'(\theta)}{\psi'(1)} \right]&= \dot{\phi}_S - \xi \dot{\theta} \frac{\psi''(\theta)}{\psi'(1)}\\
&= - \eta_2 \phi_S + \eta_1 \frac{\pi_S}{\pi}\phi_D - \beta \phi_I \frac{\psi''(\theta)}{\psi'(\theta)} \left( \phi_S - \xi \frac{\psi'(\theta)}{\psi'(1)}\right)\\
&= - \eta_2 \phi_S + \eta_1 \pi\frac{\psi'(\theta)}{\psi'(1)} - \beta \phi_I \frac{\psi''(\theta)}{\psi'(\theta)} \left( \phi_S - \xi \frac{\psi'(\theta)}{\psi'(1)}\right) + o(1)
\end{align*}
where we use the fact that $\pi_S=\phi_D\psi'(\theta)/\psi'(1)$ and $\phi_D = \pi + o(1)$ and that $\psi'(\theta)/\psi'(1) \leq 1$ (since $\psi$ is a convex function and $\theta\leq 1$).  Since $\eta_1\pi=\eta_2\xi$, we have
\begin{align*}
\diff{}{t} \left[\phi_S - \xi \frac{\psi'(\theta)}{\psi'(1)} \right]&= - \eta_2 \left(\phi_S - \xi\frac{\psi'(\theta)}{\psi'(1)}\right) - \beta \phi_I \frac{\psi''(\theta)}{\psi'(\theta)} \left( \phi_S - \xi \frac{\psi'(\theta)}{\psi'(1)}\right) + o(1)\\
&= - \left(\eta_2 + \beta \phi_I \frac{\psi''(\theta)}{\psi'(\theta)} \right)\left(\phi_S - \xi\frac{\psi'(\theta)}{\psi'(1)}\right) + o(1)
\end{align*}
 Since $\psi''(\theta)/\psi'(\theta)$ is nonnegative, the coefficient of $\phi_S - \xi\psi'(\theta)/\psi'(1)$ in the final expression is at least $\eta_2$, so we conclude that $\phi_S = \xi \psi'(\theta)/\psi'(1) + o(1)$.

 \item $\pi_I = \pi \zeta + o(1)$ \qquad $\xi_I = \xi \zeta + o(1)$.

We first define $q = (-\phi_D + \pi \theta) \psi'(\theta)/\psi'(1)$.  Since $\theta = 1+o(1)$ and $\phi_D = \pi + o(1)$, we have $-\phi_D + \pi \theta = o(1)$.  Since $0 \leq \psi'(\theta) \leq \psi'(1)$, we have $q = o(1)$.  We will show that $\pi_I - \pi \zeta - q = o(1)$
\begin{align*}
\diff{}{t} (\pi_I - \pi \zeta -q) &= - \dot{\pi}_S - \gamma \pi_I - \pi \dot{\zeta} -\dot{q}\\
                 &= -\diff{}{t} \left(\phi_D \frac{\psi'(\theta)}{\psi'(1)}\right) - \gamma \pi_I - \pi \left( - \diff{}{t} \left(\theta \frac{\psi'(\theta)}{\psi'(1)} \right) - \gamma \zeta \right) - \diff{}{t} \left((-\phi_D+\pi\theta)\frac{\psi'(\theta)}{\psi'(1)}\right)\\
&= - \gamma (\pi_I + \pi \zeta - q) + o(1)
\end{align*}
So $\pi_I - \pi \zeta - q =o(1)$, and $\pi_I = \pi \zeta + o(1)$.  A similar proof shows that $\xi_I = \xi \zeta+o(1)$.

\item $\phi_I = \xi \theta \zeta + o(1)$

We will actually show that $\phi_I + \phi_S = \xi\theta\zeta + \xi\theta^2\psi'(\theta)/\psi'(1) + o(1)$.  
Once this result is shown, we use the fact that $\phi_S = \xi\psi'(\theta)/\psi'(1) + o(1)=\xi\theta^2\psi'(\theta)/\psi'(1)+o(1)$ to get our final result.  
We begin by taking the derivative of $\phi_I +\phi_S - \xi [\theta \zeta + \theta^2\psi'(\theta)/\psi'(1)] $
\[
\diff{}{t} \left(\phi_I +\phi_S - \xi \left[\theta \zeta + \theta^2\frac{\psi'(\theta)}{\psi'(1)}\right] \right) = \dot{\phi}_I + \dot{\phi}_S - \xi \left[ \dot{\theta}\zeta + \theta\dot{\zeta} +\dot{\theta}\theta^2\frac{\psi''(\theta)}{\psi'(1)}
+ 2\dot{\theta} \theta\frac{\psi'(\theta)}{\psi'(1)} 
\right]
\]
After some manipulations (see Appendix) we have

\[
\diff{}{t} \left(\phi_I +\phi_S - \xi \left[\theta \zeta + \theta^2\frac{\psi'(\theta)}{\psi'(1)}\right] \right)  
= - (\eta_2+\gamma) \left(\phi_I +\phi_S - \xi \left[\theta \zeta + \theta^2\frac{\psi'(\theta)}{\psi'(1)}\right] \right)  + o(1)
\]
So $\phi_I+\phi_S-\xi[\theta\zeta + \theta^2\psi(\theta)/\psi'(1)] = o(1)$.  Because $\theta=1+o(1)$ and $\phi_S=\xi\psi'(\theta)/\psi'(1)+o(1)$ it follows then that 
$\phi_I = \xi \theta \zeta + o(1)$.

\item $\dot{\theta} = -\tilde{\beta}\theta \zeta + \beta o(1)$.

This step is trivial.  Since $\phi_I = \xi\theta\zeta + o(1)$, we have 
\begin{align*}
\dot{\theta} &= -\beta \phi_I \\
&= - \tilde{\beta} \theta \zeta + \beta o(1)
\end{align*}

\item $\zeta = 1 - \theta\frac{\psi'(\theta)}{\psi'(1)} + (\gamma/\tilde{\beta})\ln \theta +t o(1)$ 

We make the observation that $\dot{\zeta} = -\diff{}{t} [\theta\psi'(\theta)/\psi'(1)] - \gamma \zeta$.   
We have
\begin{align*}
\dot{\zeta} - \diff{}{t} \left[ 1 - \theta\frac{\psi'(\theta)}{\psi'(1)} + (\gamma/\tilde{\beta})\ln \theta\right] &= -\diff{}{t} \left[\theta\frac{\psi'(\theta)}{\psi'(1)}\right] - \gamma \zeta + \diff{}{t} \left[\theta\frac{\psi'(\theta)}{\psi'(1)}\right] - \frac{\gamma \dot{\theta}}{\tilde{\beta} \theta}\\
&= -\gamma \zeta - \frac{\gamma\dot{\theta}}{\tilde{\beta}\theta}
\end{align*}
From our previous result we have $\zeta = - [\dot{\theta}+\beta o(1)]/\tilde{\beta}\theta$.  Substituting this in we have
\begin{align*}
\dot{\zeta} - \diff{}{t} \left[ 1 - \theta\frac{\psi'(\theta)}{\psi'(1)} + (\gamma/\tilde{\beta})\ln \theta\right] &= - \frac{\gamma\dot{\theta}}{\tilde{\beta}\theta} - \frac{\gamma \beta o(1)}{\tilde{\beta}\theta} - \frac{\gamma \dot{\theta}}{\tilde{\beta}\theta}\\
= o(1)
\end{align*}
Our result follows immediately.
\end{enumerate}
So the DC model reduces to the actual degree formulation of the MFSH model.  Because the expected degree formulation contains the actual degree formulation as a special case, it suffices now to prove that the expected degree formulation converges to the MA model in the appropriate limit.

\subsection{MFSH to MA}
\label{sec:MFSH2MArigor}
Our proof that the MFSH model converges to the MA model does not require that $\ave{K} \to \infty$, but it does require $\ave{K^4}/\ave{K}^4 \to 1$.  However, in order to conclude that the other models approach the MA model, we need the conditions of the previous section to hold as well.

We first make the observation that if $\ave{K^4}/\ave{K}^4 \to 1$, then $\ave{K^2}/\ave{K}^2 \to 1$ as well.  By Jensen's inequality, we have $\ave{K}^2 \leq \ave{K^2} \leq \sqrt{\ave{K^4}}$.  So $1=\ave{K}^2/\ave{K}^2 \leq \ave{K^2}/\ave{K}^2 \leq \sqrt{\ave{K^4}}/\ave{K}^2 = \sqrt{\ave{K^4}/\ave{K}^4}\to 1$.  Sp $\ave{K^2}/\ave{K}^2$ is bounded from below by $1$ and from above by something converging to $1$.


We first prove two lemmas.  We define $s_1$ and $s_2$ by
\begin{align*}
s_1 &=\frac{1}{\ave{K}}\frac{\Psi''(\Theta)}{\Psi'(1)} - S\\
s_2 &=\frac{1}{\ave{K}} \Psi'(\Theta)-S
\end{align*}
We show $s_1$ and $s_2$ go to zero.  We have
\begin{align*}
\lim_{n\to\infty}|s_1| &= \lim_{n\to\infty}\left | \frac{\Psi''(\Theta)}{\ave{K}^2} - S \right|\\
&= \lim_{n\to\infty}\left| \frac{\int_0^\infty (e^{-\kappa(1-\Theta)} - S)(\kappa^2-\ave{K^2})\, \mathrm{d}\kappa}{\ave{K}^2} \right|
\end{align*}
(see Appendix for details).
We now use the Cauchy-Schwarz inequality to bound this.  We have
\begin{align*}
\left| \frac{\int_0^\infty (e^{-\kappa(1-\Theta)} - S)(\kappa^2-\ave{K^2})\, \mathrm{d}\kappa}{\ave{K}^2} \right| &\leq \frac{|\int_0^\infty (e^{-\kappa(1-\Theta)} - S)^2 \rho(\kappa) \, \mathrm{d}\kappa|^{1/2} |\int_0^\infty (k^2-\ave{K^2})^2 \, \mathrm{d}\kappa |^{1/2}}{\ave{K}^2}\\
&\leq \frac{|\int_0^\infty (\kappa^2-\ave{K^2})^2 \, \mathrm{d}\kappa |^{1/2}}{\ave{K}^2}
\end{align*}
where we use the fact that $|\exp[-\kappa(1-\Theta)]-S|\leq 1$ to show that the first term in the numerator of the first equation is at most $1$.  All that remains is to expand the numerator and bound it.
\begin{align*}
\frac{|\int_0^\infty (\kappa^2-\ave{K^2})^2 \, \mathrm{d}\kappa |^{1/2}}{\ave{K}^2} &= \frac{|\int_0^\infty (\kappa^4-2\kappa^2\ave{K^2}+\ave{K^2}^2) \, \mathrm{d}\kappa |^{1/2}}{\ave{K}^2}\\
&= \left|\frac{\ave{K^4}-\ave{K^2}^2}{\ave{K}^4}\right|^{1/2}
\end{align*}
But our assumption is that as $n\to\infty$,  both $\ave{K^4}/\ave{K}^4$ and $\ave{K^2}/\ave{K}^2$ go to $1$.  Thus this goes to zero as $n \to \infty$.  This completes the proof that $s_1 \to 0$.

To show that $s_2 \to 0$ is similar.  
\begin{align*}
|s_2| &=\left|\frac{1}{\ave{K}} \Psi'(\Theta)-S\right|\\
&= \left| \frac{\int_0^\infty (e^{-\kappa(1-\Theta)} - S)(\kappa-\ave{K}) \rho(\kappa) \, \mathrm{d}\kappa}{\ave{K}} \right|\\
& \leq \frac{| \int_0^\infty (e^{-\kappa(1-\Theta)} - S) \rho(\kappa) \, \mathrm{d}\kappa|^{1/2} |\int_0^\infty (\kappa-\ave{K}) \rho(\kappa) \, \mathrm{d}\kappa |^{1/2} }{\ave{K}}\\
&\leq \frac{|\ave{K^2}-\ave{K}^2|^{1/2}}{\ave{K}}
\end{align*}
and as $n\to \infty$, this tends to $0$ as well.

We have our bounds on $s_1$ and $s_2$, so we are now ready to complete the proof that the Mean Field Social Heterogeneity model converges to the Mass Action model as $n \to \infty$.

We define $\Pi_I = -\dot{\Theta}/\beta =  1- \Psi'(\Theta)/\Psi'(1) - \gamma(1-\Theta)/\beta$.  In~\cite{miller:ebcm_overview} we saw that this represents the probability that a new neighbor is infected.  We expect $\Pi_I$ to be very close to $I$ and set $y = \Pi_I - I$.  Then
\begin{align*}
\dot{y} &= \dot{\Pi}_I - \dot{I}\\
 &= -\dot{\Theta} \frac{\Psi''(\Theta)}{\Psi'(1)} + \gamma \dot{\Theta}/\beta - [-\dot{S} - \dot{R}] \\
&=  -\dot{\Theta} \frac{\Psi''(\Theta)}{\Psi'(1)} + \gamma \dot{\Theta}/\beta - [-\dot{\Theta}\Psi'(\Theta) - \gamma I]\\
&= \beta \Pi_I \frac{\Psi''(\Theta)}{\Psi'(1)} - \gamma \Pi_I - \beta \Pi_I \Psi'(\Theta) + \gamma I\\
&= \beta (I+y) \frac{\Psi''(\Theta)}{\Psi'(1)} - \gamma y - \beta (I+y) \Psi'(\Theta)\\
&= \beta\ave{K} (I+y)[(S+s_1) - (S+s_2)] - \gamma y\\
&= \hat{\beta} (I+y) (s_1-s_2) - \gamma y
\end{align*}
We know that $I+y$ is at most $1$, and $s_1$ and $s_2$ both tend to zero as $n$ increases, so we have $\dot{y}$ approaches $-\gamma y$ as $n \to \infty$.  Note also that at early time $\Pi_I$ and $I$ are both very close to $0$ so $y$ begins as a very small number.  It is straightforward to conclude that $y$ can never grow larger than the maximum value of $s_1$ which tends to zero as $n \to \infty$.  Consequently, $y \to 0$ as $n \to \infty$.

We finally have
\begin{align*}
\dot{S} &= -\beta \Pi_I \Psi'(\Theta)\\
           &= -\hat{\beta} \Pi_I (S+s_2) \\
           &= -\hat{\beta} I S - \hat{\beta}yS - \hat{\beta}ys_2
\end{align*}
Since $y$ and $s_2$ both tend to $0$ as $n \to \infty$, we have $\dot{S} = -\hat{\beta}IS$.  This means that we have the Mass Action model in the limit.

\section{Discussion}
We have shown the relations between the various edge-based compartmental models for infectious disease spread derived in~\cite{miller:ebcm_overview}.  We used this to develop a flow chart (figure~\ref{fig:flowchart}) which can guide the choice of appropriate model for a given population.

We note that while the edge-based compartmental models allow us to capture effects that were previously inaccessible through analytic techniques, there are still many effects that are not captured by the models considered here.  Our flow chart does not address these.  It is always prudent to consider the disease and population to ensure that the assumptions of our models are not strongly violated.  A number of adaptations of the mass action SIR models exist for populations in which the disease has multiple stages, or the population has important substructures.  In our next paper~\cite{miller:ebcm_structure} we show that the general edge-based compartmental modeling approach can be used to accomodate many of these effects.

There is one interesting open question which we call attention to.  We showed that if $\ave{K} \to \infty$, \ $\ave{K^4}/\ave{K}^4 \to 1$ with $\beta \ave{K}$ fixed, the models converge to the Mass Action model.  However, there are many cases where $\ave{K^4}/\ave{K}^4$ approaches some other value, and our calculations appear to converge.  It would be interesting to identify what the relevant reduced equations are in this limit.

\section*{Acknowledgments}
JCM was supported by 1) the RAPIDD program of the Science and Technology Directorate, Department of Homeland Security and the Fogarty International Center, National Institutes of Health and 2) the Center for Communicable Disease Dynamics, Department of Epidemiology, Harvard School of Public Health under Award Number U54GM088558 from the National Institute Of General Medical Sciences.  EMV was supported by NIH K01 AI091440.  The content is solely the responsibility of the authors and does not necessarily represent the official views of the National Institute Of General Medical Sciences or the National Institutes of Health.

\appendix
\section{Appendix}

In this appendix we show more rigorous arguments underlying many of the claims in the main paper~\cite{miller:ebcm_hierarchy}.  We show that the various edge-based compartmental models presented in~\cite{miller:ebcm_overview} reduce to one another in appropriate limits.

We give one preliminary result that will be used several times.  For constant $c>0$, if
\[
\dot{x} = f(t) - cx
\]
then so long as $|x(0)|\leq \sup_t|f(t)|/c$, then for all time $|x(t)| \leq \sup_t f(t)/c$.   To see this, simply note that $\dot{x} <0$ if $x>\sup |f|/c$, and $\dot{x}>0$ if $x<-\sup |f|/c$.  From this fact we conclude that $x$ can never increase past $\sup |f|/c$ or decrease below $-\sup |f|/c$.  More generally, if $c$ is not constant, but bounded away from $0$, then $|x(t)| \leq \sup_t f(t) / \inf_t c(t)$.

\section{Limiting cases of the DVD model}
We begin by considering limiting cases of the DVD model.  
The equations of the DVD model are
\begin{align*}
\dot{\Theta} &= -\beta\Theta + \beta\frac{\Psi'(\Theta)}{\Psi'(1)} + \gamma(1- \Theta) + \eta\left(1- \Theta - \frac{\beta}{\gamma} \Pi_R\right) \, ,\\
\dot{\Pi}_R &= \gamma \Pi_I \, ,\qquad \Pi_S = \Psi'(\Theta)/\Psi'(1) \, ,\qquad \Pi_I = 1 - \Pi_S - \Pi_R \, , \\
\dot{R} &= \gamma I \, ,\quad\qquad S = \Psi(\Theta) \, ,\quad\qquad I=1-S-R \, .  
\end{align*}

\subsection{Convergence of DVD to MFSH if $\eta/\beta$ is large}

For a heuristic argument that large values of $\eta$ lead to the MFSH model, it is easier if we also assume that $\eta/\gamma \gg 1$.  We do this initially, and then show a more rigorous proof without this additional assumption.

The variables $\Theta$, $\Psi'(\Theta)/\Psi'(1)$ and $1-\Theta$ are all between $0$ and $1$.  We write $\Theta = \Theta_0 + \Theta_1/\eta + \cdots$ and use similar expressions for $\Pi_S$, $\Pi_I$, and $\Pi_R$.  Since $\Theta$, $\Psi'(\Theta)/\Psi'(1)$, and $1-\Theta$ are all between $0$ and $1$, and $\beta$ and $\gamma$ are small compared to $\eta$, the first terms of the $\dot{\Theta}$ equation are negligible at order $\eta$.  At order $\eta$ the equation becomes
\[
0 = \eta\left(1-\Theta_0 - \frac{\beta}{\gamma} \Pi_{R,0}\right)
\]
And so at leading order $\Theta_0 = 1-\beta\Pi_{R,0}/\gamma$.  Looking at the equation for $\dot{\Pi}_R$ we find
$\dot{\Pi}_{R,0} = \gamma \Pi_{I,0}$ and so
\begin{align*}
\dot{\Theta}_0 &= - \frac{\beta}{\gamma} \dot{\Pi}_{R,0}\\
&= -\beta \Pi_{I,0}\\
&=-\beta\left(1-\Pi_{S,0}-\Pi_{R,0}\right)\\
&= -\beta - \beta \frac{\Psi'(\Theta_0)}{\Psi'(1)} + \gamma (\Theta_0-1)
\end{align*}
This is the equation governing the evolution of $\Theta$ in the MFSH model.  The equations for $S$, $I$, and $R$ are the same in both models.  
So if $\eta/\beta \gg 1$ and $\eta/\gamma \gg 1$, the DVD model reduces to the MFSH system.

To be more rigorous and eliminate the assumption $\eta/\gamma \gg 1$, we should show that the value of the error made by using the MFSH equations cannot become large.  To do this, we set $\Omega = \Theta-1+\beta\Pi_R/\gamma$, so $\Theta = \Omega + 1 - \beta \Pi_R/\gamma$.  From the $\dot{\Theta}$ equation we find
\begin{align*}
\dot{\Omega} - \frac{\beta}{\gamma} \dot{\Pi}_R &= -\beta \Theta + \beta \frac{\Psi'(\Theta)}{\Psi'(1)} + \eta\left(1-\Theta-\frac{\beta}{\gamma}\Pi_R\right)\\
\dot{\Omega} - \beta\Pi_I &= -\beta \Theta + \beta\frac{\Psi'(\Theta)}{\Psi'(1)} - \eta \Omega\\
\dot{\Omega} - \beta\left(1-\frac{\Psi'(\Theta)}{\Psi'(1)} - \Pi_R\right) &= -\beta \Theta  + \beta\frac{\Psi'(\Theta)}{\Psi'(1)} - \eta \Omega\\
\dot{\Omega} &= -\beta(\Theta-1 + \Pi_R) - \eta \Omega
\end{align*}
The value of $|\Omega|$ can never be larger than the maximum value of $|\beta(\Theta-1+\Pi_R)|$.  Our assumption is that $\beta$ is small compared to $\eta$, so the error in $\Theta$ remains small.  Thus $\Theta$ is approximately $1-\beta\Pi_R/\gamma$, and the equations we derived above will hold.

\subsection{Convergence of DVD to MP if $\eta(t-t_0)$ is small}
When $\eta$ is small, the term relating to edge swapping becomes small in the $\dot{\Theta}$ equation.  Here we write $\Theta= \Theta_0 + \eta \Theta_1+\cdots$ and similarly expand the other variables.  The $\dot{\Theta}_0$ equation immediately matches the equation of the MP model.  
As long as $\eta \Theta_1$ is small, the solution is closely approximated by just solving for $\Theta_0$.  We find that
\[
\dot{\Theta}_1 = \left(- \beta + \beta \frac{\Psi''(\Theta_0)}{\Psi'(1)} + \gamma\right)\Theta_1 + \left(1-\Theta_0 - \frac{\beta}{\gamma} \Pi_{R,0}\right)
\]
Since $\Theta_1$ is initially $0$, we can see that it takes time for $\eta\Theta_1$ to grow to an appreciable size (and the time is longer for smaller $\eta$).  We have not been able to find a similar bound for the error as in the MFSH case, and this is because if the epidemic lasts sufficiently long, the creation and deletion of edges will rearrange the contact structure during the course of the epidemic, and so the error terms can become significant.

A more precise bound can be found by asking ``how many edges have been created or broken with nodes that were no longer susceptible?''  As long as the edges being changed are connected to susceptible individuals, there is no effect on the spread of the epidemic, but if a node creates a new edge to a recovered rather than susceptible individual, this does alter the spread.  Early in the epidemic $\Pi_I+\Pi_R$ grows exponentially and may be taken to be $C e^{rt}$ for some $r$ depending on the parameters of the disease and population.  The rate that edges with these nodes are formed or broken is proportional to $\eta (\Pi_I + \Pi_R)$.  Integrating this over time we have $\int_{-\infty}^t \eta C e^{r\tau} \, \mathrm{d}\tau = \eta Ce^{rt}/r = \eta(\Pi_I+\Pi_R)/r$.  So a measure of the quality of the approximation is whether or not $\eta(\Pi_I+\Pi_R)/r$ is large.  This captures the fact that if the edge changeover rate is slow ($\eta$ small), or the timescale of the epidemic is short ($r$ large), or the proportion of the population affected by the disease is small ($\Pi_I+\Pi_R$ small), then the impact of edge creation and deletion is not significant.

\section{Limiting cases of the DFD model}
The equations for the DFD model are
\begin{align*}
\dot{\theta} &= -\beta \phi_I \, ,\\
\dot{\phi}_S &= - \beta \phi_I \phi_S\frac{\psi''(\theta)}{\psi'(\theta)}  + \eta\theta\pi_S - \eta \phi_S\, ,  \\
\dot{\phi}_I &= \beta \phi_I \phi_S\frac{\psi''(\theta)}{\psi'(\theta)} + \eta \theta \pi_I - (\beta  + \gamma + \eta) \phi_I \, ,  \\
\dot{\pi}_R &= \gamma \pi_I \, , \qquad\qquad \pi_S = \frac{\theta\psi'(\theta)}{\psi'(1)} \, ,\qquad\qquad 
\pi_I = 1-\pi_R-\pi_S\, ,  \\ 
\dot{R} &= \gamma I \, ,\qquad\qquad S(t) = \psi(\theta) \, ,\qquad\qquad I(t) = 1-S-R \, . 
\end{align*}

\subsection{Convergence of DFD to MFSH if $\eta/\beta$ is large}

As before the simple argument is easier if we also assume that $\eta/\gamma$ is large. 
Expanding all our variables in terms of $1/\eta$, we see that at leading order $\phi_{I,0} = \theta \pi_{I,0}$.  We use this observation to reach two conclusions.  First
\begin{align*}
\dot{\pi}_{R,0} &= \gamma \pi_{I,0}\\
&= \gamma \frac{\phi_{I,0}}{\theta}\\
&= -\frac{\gamma}{\beta} \frac{\dot{\theta}}{\theta}
\end{align*}
so $\dot{\pi}_{R,0} = -\gamma [\ln \theta]/\beta$.  We further find
\begin{align*}
\dot{\theta}_0 &= -\beta \theta \pi_{I,0}\\
    & =  - \beta \theta (1-\pi_{S,0}-\pi_{R,0})\\
&= -\beta \theta\left(1-\frac{\theta\psi'(\theta)}{\psi'(1)} +\frac{\gamma \ln \theta}{\beta}\right)
\end{align*}
 and so at leading order in $1/\eta$, $\theta$ solves the MFSH equations.  

We can make this more rigorous as we did for the DVD to MFSH case, and we find that we only require $\eta/\beta$ to be large.\marginpar{check this}

\subsection{Convergence of DFD to CM if $\eta(t-t_0)$ is small}
The argument here is effectively the same as the DVD to MP case.  We just make the observation that ignoring the $\eta$ term, we find that $\phi_S=\psi'(\theta)/\psi'(1)$, so we are able to show that $\phi_I$ also takes the same solution at leading order as in the CM case.  Thus $\theta$ solves the same equation as in the CM case and the argument is done.

Again we can find a more precise condition that if $\eta(\pi_I+\pi_R)/r$ is small where $r$ is the exponential growth rate of the epidemic, then we can neglect the impact of edge swapping.

\section{Limiting cases of the DC model}

The equations for the DC model are
\begin{align*}
\dot{\theta} &= -\beta \phi_I\, , \\
\dot{\phi}_S &= - \beta \phi_I \phi_S \frac{\psi''(\theta)}{\psi'(\theta)} + \eta_1 \frac{\pi_S}{\pi}\phi_D - \eta_2\phi_S\, , \\
\dot{\phi}_I &= \beta \phi_I\phi_S\frac{\psi''(\theta)}{\psi'(\theta)} + \eta_1 \frac{\pi_I}{\pi} \phi_D - (\eta_2+\beta+\gamma)\phi_I \, , \\
\dot{\phi}_D &= \eta_2(\theta-\phi_D) - \eta_1 \phi_D\, , \\
\dot{\xi}_R &= -\eta_2 \xi_R + \eta_1 \pi_R +\gamma \xi_I \, , \qquad \xi_S = (\theta-\phi_D)\frac{\psi'(\theta)}{\psi'(1)} \, , \qquad \xi_I = \xi- \xi_S - \xi_R\, , \\
\dot{\pi}_R &= \eta_2 \xi_R-\eta_1\pi_R + \gamma \pi_I\, , \qquad \pi_S = \phi_D \frac{\psi'(\theta)}{\psi'(1)}  \, , \qquad \pi_I = \pi - \pi_S - \pi_R\, , \\
\xi &= \frac{\eta_1}{\eta_1+\eta_2} \, , \qquad \pi = \frac{\eta_2}{\eta_1+\eta_2}\, , \\
\dot{R} &= \gamma I \, , \qquad\qquad S = \psi(\theta) \, , \qquad\qquad I = 1-S-R \, .
\end{align*}
\subsection{Convergence of DC to DFD as $\pi \to 0$}
Since $\pi = \eta_2/(\eta_1+\eta_2)$, we know that as $\pi \to 0$, we must have $\eta_2/\eta_1$ to $0$, and in fact $\pi - \eta_2/\eta_1 = -\eta_2^2/(\eta_1^2+\eta_1\eta_2) \ll \pi$.  So $\pi$ is well-approximated by $\eta_2/\eta_1$.  Using this and the equation for $\dot{\phi}_D$, it is possible to show that $\phi_D = \pi \theta$ at leading order.  

From this we quickly see that $\pi_S = \pi \theta \psi'(\theta)/\psi'(1)= \pi \xi_S + \order(\pi^2)$.  A little more effort shows that $\pi_I = \pi \xi_I + \order(\pi^2)$ and $\pi_R = \pi \xi_R + \order(\pi^2)$.  We then find at leading order
\begin{align*}
\dot{\phi}_S &= -\beta \phi_I \phi_S \frac{\psi''(\theta)}{\psi'(\theta)} + \eta_2 \xi_S \theta  - \eta_2 \phi_S\\
\dot{\phi}_I &= \beta \phi_I \phi_S \frac{\psi''(\theta)}{\psi'(\theta)} + \eta_2 \xi_I \theta - (\eta_2+\beta+\gamma)\phi_I
\end{align*}
where we have used the fact that $\eta_1 \phi_D = \eta_2 \theta$ at leading order.  

We can also show that the $\xi$ variables in the DC model follow the same equations as the $\pi$ variables in the DFD model.  Letting $\eta_2$ play the role of $\eta$ n the DFD model,  our DC equations reduce to the DFD equations.

\subsection{Convergence of DC to DVD as $\pi \to 1$}

Consider the variable-degree model with a given $\rho(\kappa)$.  We want to build a dynamic fixed-degree network that mimics this.  Epidemics on this network may be modeled using the dormant contact equations, but the network will behave very much like a variable-degree network. Our approach is as follows: we will take some $L \gg 1$ and consider the dormant edge model where each node that would have expected degree $\kappa$ has $\floor{L\kappa}$ stubs (where $\floor{\cdot}$ is the integer part) with the edge swapping occurring in such a way that only a fraction $1/L$ of the stubs are active.  This is done by having existing edges break at rate $\eta_2=\eta$ and dormant stubs finding neighbors at rate $\eta_1=\eta/L$.  The leading order terms (in $L$) in the equations of the dormant edge model will be the same as the variable-degree model. 

As $L \to \infty$, we anticipate that $\pi =1 + \order(1/L)$ and $\phi_D = 1 + \order(1/L)$. We also expect that $\xi$ variables are all $\order(1/L)$ and negligible in comparison to all other terms.  
We set $P(k,L) = \int_{k/L}^{(k+1)/L} \rho(\kappa) \, \mathrm{d}\kappa$ which corresponds to taking all the weight of $\kappa$ and assigning it to $\floor{L\kappa}$.  Thus a node that would have expected degree $\kappa$  with $k \leq \kappa L < k+1$ instead gets $k$ stubs with $k$ a close approximation to $L\kappa$.  Define $\psi(x,L) = \sum_k x^k P(k,L)$.

We set
\[
\Psi(x) = \int_0^\infty e^{-\kappa(1-x)}\rho(\kappa) \mathrm{d}\kappa
\]
Through the substitutions
\begin{gather*}
\theta^L = e^{\Theta-1}\\
\phi_S = \frac{\Phi_S}{L} \qquad\qquad \phi_I = \frac{\Phi_I}{L}\\
\pi_S = \Pi_S\pi\qquad\qquad
\pi_I = \Pi_I\pi\qquad\qquad
\pi_R = \Pi_R\pi\\
\eta_2/\eta_1 = L \qquad\qquad \eta_2 = \eta
\end{gather*}
and starting from the fixed-degree equations we arrive at the variable-degree equations in the $L \to \infty$ limit.  To be fully rigorous, we would define $\Theta$, $\Phi_S$, $\Pi_S$, etc.\ to be the limits as $L \to \infty$.

To show this, we first show that $\psi(\theta,L) = \Psi(\Theta) +\order(1/L)$.  Take $L \gg 1$ to be given.  We have
\begin{align*}
\psi(\theta,L) &= \sum_{k=0}^\infty \theta^k P(k,L) \\
&= \sum_{k=0}^\infty \theta^{k} \left[\int_{k/L}^{(k+1)/L}\rho(\kappa) \, \mathrm{d}\kappa\right] \\
&=\sum_{k=0}^\infty  \left[\int_{k/L}^{(k+1)/L}\theta^{L\kappa}\rho(\kappa) \, \mathrm{d}\kappa\right] + \order(1/L)\\
&= \int_0^\infty \theta^{L\kappa} \rho(\kappa) \, \mathrm{d}\kappa  + \order(1/L)\\
&= \int_0^\infty e^{-\kappa(1-\Theta)}\rho(\kappa) \, \mathrm{d}\kappa  + \order(1/L)\\
&=\Psi(\Theta)  + \order(1/L)
\end{align*}
Note that $\mathrm{d}/\mathrm{d}\Theta = (\theta/L) \mathrm{d}/\mathrm{d}\theta$, and when $L$ is large, $\theta = 1 + \order(1/L)$.  
From this it follows that for $n=1,2$, \ $\psi^{(n)}(\theta,L) = L^n\Psi^{(n)}(\Theta) + \order(L^{n-1})$.
We now take the DC equations and substitute these values for the variables.   Holding on to just leading order terms in the equations and observing that $\Phi_S = \Psi'(\Theta)/\Psi'(1)$, we see that the variables scale as anticipated, and the equations reduce to the dynamic variable-degree equations at leading order.

\subsection{Convergence of DC to MFSH as $\eta_1/\beta \to \infty$ with $\eta_2/\eta_1$ fixed}

The argument for this is much the same as in the DFD to MFSH and DVD to MFSH.  One modification is needed.  We note that many edges in the DC model will not be active at any given time: the proportion of time an edge is active is $\xi$.  The appropriate transmission rate in the MFSH model is thus $\hat{\beta} = \xi \beta$.  Taking the leading order equations in $1/\eta_1$, and following the same basic approach as before, we arrice at the actual degree formulation of the MFSH equations.

\section{Convergence of MFSH to MA if $\ave{K^4}/\ave{K}^4 \to 1$}

We first make the observation that if $\ave{K^4}/\ave{K}^4 \to 1$, then $\ave{K^2}/\ave{K}^2 \to 1$ as well.  By Jensen's inequality, we have $\ave{K}^2 \leq \ave{K^2} \leq \sqrt{\ave{K^4}}$.  So $1=\ave{K}^2/\ave{K}^2 \leq \ave{K^2}/\ave{K}^2 \leq \sqrt{\ave{K^4}}/\ave{K}^2 = \sqrt{\ave{K^4}/\ave{K}^4}\to 1$.  Sp $\ave{K^2}/\ave{K}^2$ is bounded from below by $1$ and from above by something converging to $1$.

Let us assume we have a sequence of populations and diseases (indexed by $n$) such that $\gamma$ and $\beta \ave{K}=\hat{\beta}$ are held fixed.  Further assume that $\ave{K^4}/\ave{K}^4 \to 1$ as $n \to \infty$.  We will prove that the solution of the MFSH model converges to the solution of the Mass Action model.

We first prove two lemmas.  We define $s_1$ and $s_2$ by
\begin{align*}
s_1 &=\frac{1}{\ave{K}}\frac{\Psi''(\Theta)}{\Psi'(1)} - S\\
s_2 &=\frac{1}{\ave{K}} \Psi'(\Theta)-S
\end{align*}
We show $s_1$ and $s_2$ go to zero.  We have
\begin{align*}
\lim_{n\to\infty}|s_1| &= \lim_{n\to\infty}\left | \frac{\Psi''(\Theta)}{\ave{K}^2} - S \right|\\
&= \lim_{n\to\infty}\left |\frac{\int_0^\infty \kappa^2 e^{-\kappa(1-\Theta)} \rho(\kappa) \, \mathrm{d}\kappa}{\ave{K}^2} - S\right|\\
&= \lim_{n\to\infty}\left| \frac{\int_0^\infty \kappa^2 e^{-\kappa(1-\Theta)} \rho(\kappa) \, \mathrm{d}\kappa}{\ave{K}^2} - S \frac{\int_0^\infty \kappa^2 \rho(\kappa) \, \mathrm{d}\kappa}{\ave{K^2}} \right|\\
&= \lim_{n\to\infty}\left| \frac{\int_0^\infty \kappa^2 e^{-\kappa(1-\Theta)} \rho(\kappa) \, \mathrm{d}\kappa}{\ave{K}^2} - S \frac{\int_0^\infty \kappa^2 \rho(\kappa) \, \mathrm{d}\kappa}{\ave{K}^2} \right |\\
&= \lim_{n\to\infty}\left| \frac{\int_0^\infty (e^{-\kappa(1-\Theta)} - S)(\kappa^2-\ave{K^2})\, \mathrm{d}\kappa}{\ave{K}^2} \right|
\end{align*}
where in the next to last step we use $\ave{K^2} \to \ave{K}^2$, and in the last step we note $\int (\exp[-\kappa(1-\Theta)]-S)\ave{K^2}\mathrm{d}\kappa =0$.  We now use the Cauchy-Schwarz inequality to bound this.  We have
\begin{align*}
\left| \frac{\int_0^\infty (e^{-\kappa(1-\Theta)} - S)(\kappa^2-\ave{K^2})\, \mathrm{d}\kappa}{\ave{K}^2} \right| &\leq \frac{|\int_0^\infty (e^{-\kappa(1-\Theta)} - S)^2 \rho(\kappa) \, \mathrm{d}\kappa|^{1/2} |\int_0^\infty (k^2-\ave{K^2})^2 \, \mathrm{d}\kappa |^{1/2}}{\ave{K}^2}\\
&\leq \frac{|\int_0^\infty (\kappa^2-\ave{K^2})^2 \, \mathrm{d}\kappa |^{1/2}}{\ave{K}^2}
\end{align*}
where we use the fact that $|\exp[-\kappa(1-\Theta)]-S|\leq 1$ to show that the first term in the numerator of the first equation is at most $1$.  All that remains is to expand the numerator and bound it.
\begin{align*}
\frac{|\int_0^\infty (\kappa^2-\ave{K^2})^2 \, \mathrm{d}\kappa |^{1/2}}{\ave{K}^2} &= \frac{|\int_0^\infty (\kappa^4-2\kappa^2\ave{K^2}+\ave{K^2}^2) \, \mathrm{d}\kappa |^{1/2}}{\ave{K}^2}\\
&= \left|\frac{\ave{K^4}-\ave{K^2}^2}{\ave{K}^4}\right|^{1/2}
\end{align*}
But our assumption is that as $n\to\infty$,  both $\ave{K^4}/\ave{K}^4$ and $\ave{K^2}/\ave{K}^2$ go to $1$.  Thus this goes to zero as $n \to \infty$.  This completes the proof that $s_1 \to 0$.

To show that $s_2 \to 0$ is similar.  
\begin{align*}
|s_2| &=\left|\frac{1}{\ave{K}} \Psi'(\Theta)-S\right|\\
&= \left| \frac{\int_0^\infty \kappa e^{-\kappa(1-\Theta)}\rho(\kappa) \, \mathrm{d}\kappa}{\ave{K}} - \frac{S \int_0^\infty \kappa \rho(\kappa) \, \mathrm{d}\kappa}{\ave{K}} \right |\\
&= \left| \frac{\int_0^\infty (e^{-\kappa(1-\Theta)} - S)(\kappa-\ave{K}) \rho(\kappa) \, \mathrm{d}\kappa}{\ave{K}} \right|\\
& \leq \frac{| \int_0^\infty (e^{-\kappa(1-\Theta)} - S) \rho(\kappa) \, \mathrm{d}\kappa|^{1/2} |\int_0^\infty (\kappa-\ave{K}) \rho(\kappa) \, \mathrm{d}\kappa |^{1/2} }{\ave{K}}\\
&\leq \frac{|\int_0^\infty (\kappa-\ave{K}) \rho(\kappa) \, \mathrm{d}\kappa |^{1/2} }{\ave{K}}\\
&\leq \frac{|\ave{K^2}-\ave{K}^2|^{1/2}}{\ave{K}}
\end{align*}
and as $n\to \infty$, this tends to $0$ as well.

We have our bounds on $s_1$ and $s_2$, so we are now ready to complete the proof that the Mean Field Social Heterogeneity model converges to the Mass Action model as $n \to \infty$.

In~\cite{miller:ebcm_overview}, we showed that
$\Pi_I = 1- \Psi'(\Theta)/\Psi'(1) - \gamma(1-\Theta)/\beta$.  We expect $\Pi_I$ to be very close to $I$.  Set $y = \Pi_I - I$.  Then
\begin{align*}
\dot{y} &= \dot{\Pi}_I - \dot{I}\\
 &= -\dot{\Theta} \frac{\Psi''(\Theta)}{\Psi'(1)} + \gamma \dot{\Theta}/\beta - [-\dot{S} - \dot{R}] \\
&=  -\dot{\Theta} \frac{\Psi''(\Theta)}{\Psi'(1)} + \gamma \dot{\Theta}/\beta - [-\dot{\Theta}\Psi'(\Theta) - \gamma I]\\
&= \beta \Pi_I \frac{\Psi''(\Theta)}{\Psi'(1)} - \gamma \Pi_I - \beta \Pi_I \Psi'(\Theta) + \gamma I\\
&= \beta (I+y) \frac{\Psi''(\Theta)}{\Psi'(1)} - \gamma y - \beta (I+y) \Psi'(\Theta)\\
&= \beta\ave{K} (I+y)[(S+s_1) - (S+s_2)] - \gamma y\\
&= \hat{\beta} (I+y) (s_1-s_2) - \gamma y
\end{align*}
We know that $I+y$ is at most $1$, and $s_1$ and $s_2$ both tend to zero as $n$ increases, so we have $\dot{y}$ approaches $-\gamma y$ as $n \to \infty$.  Note also that at early time $\Pi_I$ and $I$ are both very close to $0$ so $y$ begins as a very small number.  It is straightforward to conclude that $y$ can never grow larger than the maximum value of $s_1$ which tends to zero as $n \to \infty$.  Consequently, $y \to 0$ as $n \to \infty$.

We finally have
\begin{align*}
\dot{S} &= -\beta \Pi_I \Psi'(\Theta)\\
           &= -\hat{\beta} \Pi_I (S+s_2) \\
           &= -\hat{\beta} I S - \hat{\beta}yS - \hat{\beta}ys_2
\end{align*}
Since $y$ and $s_2$ both tend to $0$ as $n \to \infty$, we have $\dot{S} = -\hat{\beta}IS$.  This means that we have the Mass Action model in the limit.
 
\section{Dormant Contact model to Mean Field Social Heterogeneity model as $\ave{K} \to \infty$, \ $\beta = \tilde{\beta}/\xi$, with $\tilde{\beta}\ave{K}$, $\gamma$ constant}

We want to prove that if degrees increase, but simultaneously the infection rate decreases with $\gamma$ constant then we must arrive at the Mass Action equations.  We do this for the Dormant Contact model.  We introduce the notation $o(1)$ to denote a function whose value goes to zero as $\ave{K} \to \infty$.  We will show that neglecting small terms in the large degree limit the DC model becomes
\begin{align*}
\dot{\theta} &= -\tilde{\beta} \theta \zeta\\
\dot{R} &= \gamma I\,, \qquad S=\psi(\theta) \, , \qquad I = 1-S-R
\end{align*}
where $\zeta = 1 - \theta \psi'(\theta)/\psi'(1) + (\gamma/\tilde{\beta})\ln \theta$. 

To be more rigorous, we define $\zeta = \pi_I + \xi_I$  This is the proportion of all stubs which belong to infected nodes.  We prove a series of results:
\begin{enumerate}
\item $\theta = 1 + o(1)$.
\item $\phi_D = \pi + o(1)$.
\item $\phi_S = \xi \frac{\psi'(\theta)}{\psi'(1)} + o(1)$.
 \item $\pi_I = \pi \zeta + o(1)$ \qquad $\xi_I = \xi \zeta + o(1)$.
 \item $\phi_I = \xi \theta \zeta + o(1)$
 \item $\dot{\theta} = -\tilde{\beta}\theta \zeta  + \beta o(1)$.
\item $\zeta = 1 - \theta\frac{\psi'(\theta)}{\psi'(1)} + (\gamma/\tilde{\beta})\ln \theta + t o(1)$
 \end{enumerate}
If we drop the error terms in the final two results, then we have the equations governing the Mean Field Social Heterogeneity model in the actual degree case, with $\tilde{\beta}=\beta \xi$ playing the role of the transmission rate.  The existence of the error terms means that at earlier time, the approximation is better, and it can deviate more as time increases.  The solution converges uniformly for any $t$ less than any arbitrary chosen value $T$ as $\ave{K}$ grows.  So for large enough $\ave{K}$, the region over which the MFSH model gives an accurate approximation will include the entire period of the epidemic. 

\begin{enumerate}
\item $\theta = 1 + o(1)$.

We trivially have $1 \geq \theta \geq 1- \beta/(\beta+\gamma)$.  Because $\beta \ave{K}$ is constant, we find $\beta = o(1)$.  Thus $1 \geq \theta \geq 1 + o(1)$ (with a negative error term), and so we can conclude that $\theta = 1 + o(1)$.

\item $\phi_D = \pi + o(1)$.

We seek to prove that $\phi_D$ is approximately $\pi$.  The way we will do this is to prove instead that $\phi_D-\pi = o(1)$.   Since $\phi_D = \pi$ before the disease is introduced, the error never has a chance to be larger than $o(1)$.  We will use this approach for most of our remaining results.
\begin{align*}
\diff{}{t} (\phi_D - \pi) &= \eta_2 (\theta - \phi_D) - \eta_1 \phi_D\\
   &= \eta_2[1 + o(1)] - (\eta_1 + \eta_2) \phi_D\\
  &=  (\eta_1+\eta_2)\pi - (\eta_1+\eta_2)\phi_D + o(1)\\
  &= -(\eta_1+\eta_2)(\phi_D-\pi) + o(1)
\end{align*}
In the next to last step we used the fact that $\pi = \eta_2/(\eta_1+\eta_2)$ so $\eta_2 = (\eta_1+\eta_2)\pi$.  So $\phi_D - \pi$ is at most $o(1)/(\eta_1+\eta_2) = o(1)$.

 \item $\phi_S = \xi \frac{\psi'(\theta)}{\psi'(1)} + o(1)$.

We have
\begin{align*}
\diff{}{t} \left[\phi_S - \xi \frac{\psi'(\theta)}{\psi'(1)} \right]&= \dot{\phi}_S - \xi \dot{\theta} \frac{\psi''(\theta)}{\psi'(1)}\\
&= - \beta \phi_I \phi_S \frac{\psi''(\theta)}{\psi'(\theta)}+ \eta_1 \frac{\pi_S}{\pi}\phi_D - \eta_2\phi_S + \xi \beta \phi_I \frac{\psi''(\theta)}{\psi'(1)}\\
&= - \eta_2 \phi_S + \eta_1 \frac{\pi_S}{\pi}\phi_D - \beta \phi_I \frac{\psi''(\theta)}{\psi'(\theta)} \left( \phi_S - \xi \frac{\psi'(\theta)}{\psi'(1)}\right)\\
&= - \eta_2 \phi_S + \eta_1 \frac{\phi_D^2}{\pi}\frac{\psi'(\theta)}{\psi'(1)} - \beta \phi_I \frac{\psi''(\theta)}{\psi'(\theta)} \left( \phi_S - \xi \frac{\psi'(\theta)}{\psi'(1)}\right)\\
&= - \eta_2 \phi_S + \eta_1 \pi\frac{\psi'(\theta)}{\psi'(1)} - \beta \phi_I \frac{\psi''(\theta)}{\psi'(\theta)} \left( \phi_S - \xi \frac{\psi'(\theta)}{\psi'(1)}\right) + o(1)
\end{align*}
where we use the fact that $\phi_D = \pi + o(1)$ and that $\psi'(\theta)/\psi'(1) \leq 1$ (since $\psi$ is a convex function and $\theta\leq 1$).  Since $\eta_1\pi=\eta_2\xi$, we have
\begin{align*}
\diff{}{t} \left[\phi_S - \xi \frac{\psi'(\theta)}{\psi'(1)} \right]&= - \eta_2 \left(\phi_S - \xi\frac{\psi'(\theta)}{\psi'(1)}\right) - \beta \phi_I \frac{\psi''(\theta)}{\psi'(\theta)} \left( \phi_S - \xi \frac{\psi'(\theta)}{\psi'(1)}\right) + o(1)\\
&= - \left(\eta_2 + \beta \phi_I \frac{\psi''(\theta)}{\psi'(\theta)} \right)\left(\phi_S - \xi\frac{\psi'(\theta)}{\psi'(1)}\right) + o(1)\\
\end{align*}
 Since $\psi''(\theta)/\psi'(\theta)$ is nonnegative, the coefficient of $\phi_S - \xi\psi'(\theta)/\psi'(1)$ in the final expression is at least $\eta_2$, so we conclude that $\phi_S = \xi \psi'(\theta)/\psi'(1) + o(1)$.

 \item $\pi_I = \pi \zeta + o(1)$ \qquad $\xi_I = \xi \zeta + o(1)$.

We first define $q = (-\phi_D + \pi \theta) \psi'(\theta)/\psi'(1)$.  Since $\theta = 1+o(1)$ and $\phi_D = \pi + o(1)$, we have $-\phi_D + \pi \theta = o(1)$.  Since $0 \leq \psi'(\theta) \leq \psi'(1)$ ($\psi$ is a convex function), we have $q = o(1)$.  We will show that $\pi_I - \pi \zeta - q = o(1)$
\begin{align*}
\diff{}{t} (\pi_I - \pi \zeta -q) &= - \dot{\pi}_S - \gamma \pi_I - \pi \dot{\zeta} -\dot{q}\\
                 &= -\diff{}{t} \left(\phi_D \frac{\psi'(\theta)}{\psi'(1)}\right) - \gamma \pi_I - \pi \left( - \diff{}{t} \left(\theta \frac{\psi'(\theta)}{\psi'(1)} \right) - \gamma \zeta \right) - \diff{}{t} \left((-\phi_D+\pi\theta)\frac{\psi'(\theta)}{\psi'(1)}\right)\\
&= - \gamma (\pi_I + \pi \zeta)\\
&= - \gamma (\pi_I + \pi \zeta - q) + o(1)
\end{align*}
So $\pi_I - \pi \zeta - q =o(1)$, and $\pi_I = \pi \zeta + o(1)$.  A similar proof shows that $\xi_I = \xi \zeta+o(1)$, or we can simply use 
\begin{align*}
\xi_I &= \zeta - \pi_I \\
&= \zeta - \pi \zeta + o(1)\\
&= \xi \zeta + o(1)
\end{align*}

\item $\phi_I = \xi \theta \zeta + o(1)$

To show this, we will actually show that $\phi_I + \phi_S = \xi\theta\zeta + \xi\theta^2\psi'(\theta)/\psi'(1) + o(1)$.  In doing so we take advantage of the fact that multiplying by $\theta$ behaves very much like multiplying by $1$.   Once this result is shown, we use the fact that $\phi_S = \xi\psi'(\theta)/\psi'(1) + o(1)=\xi\theta^2\psi'(\theta)/\psi'(1)+o(1)$ to get our final result.  
We begin by taking the derivative of $\phi_I +\phi_S - \xi [\theta \zeta + \theta^2\psi'(\theta)/\psi'(1)] $
\begin{align*}
\diff{}{t} \left(\phi_I +\phi_S - \xi \left[\theta \zeta + \theta^2\frac{\psi'(\theta)}{\psi'(1)}\right] \right) &= \dot{\phi}_I + \dot{\phi}_S - \xi \left[ \dot{\theta}\zeta + \theta\dot{\zeta} +\dot{\theta}\theta^2\frac{\psi''(\theta)}{\psi'(1)}
+ 2\dot{\theta} \theta\frac{\psi'(\theta)}{\psi'(1)} 
\right]\\
&= -\eta_2 (\phi_I + \phi_S) + \eta_1 \frac{\phi_D }{\pi} (\pi_S + \pi_I) - (\beta+\gamma)\phi_I  \\
&\qquad+ \xi \beta\phi_I\left(\zeta +\theta^2\frac{\psi''(\theta)}{\psi'(1)} \right) + \xi\theta \left(\diff{}{t} \theta \frac{\psi'(\theta)}{\psi'(1)} + \gamma\zeta \right) -2\beta\theta\phi_I\frac{\psi'(\theta)}{\psi'(1)}\\
&= -\eta_2 (\phi_I + \phi_S) + \eta_1 
\left(\pi\frac{\psi'(\theta)}{\psi'(1)} + \pi\zeta\right) - (\beta+\gamma)\phi_I  \\
&\qquad+ \xi \beta\phi_I\left(\zeta +\theta^2\frac{\psi''(\theta)}{\psi'(1)} \right) + \xi\theta \left(\diff{}{t} \theta \frac{\psi'(\theta)}{\psi'(1)} + \gamma\zeta \right) -2\beta\theta\phi_I\frac{\psi'(\theta)}{\psi'(1)}+o(1)\\
&= -\eta_2 \left(\phi_I + \phi_S-\xi\frac{\psi'(\theta)}{\psi'(1)} - \xi\zeta\right) - (\beta+\gamma)\phi_I  \\
&\qquad+ \xi \beta\phi_I\left(\zeta +\theta^2\frac{\psi''(\theta)}{\psi'(1)} \right) + \xi\theta \left(\diff{}{t} \theta \frac{\psi'(\theta)}{\psi'(1)} + \gamma\zeta \right) -2\beta\theta\phi_I\frac{\psi'(\theta)}{\psi'(1)}+o(1)\\
&= -\eta_2 \left(\phi_I + \phi_S-\xi\frac{\psi'(\theta)}{\psi'(1)} - \xi\zeta\right) - \beta\phi_I -\gamma(\phi_I-\xi\theta\zeta)  \\
&\qquad+ \xi \beta\phi_I\left(\zeta +\theta^2\frac{\psi''(\theta)}{\psi'(1)} \right) + \xi\theta \left(\diff{}{t} \theta \frac{\psi'(\theta)}{\psi'(1)} \right) -2\beta\theta\phi_I\frac{\psi'(\theta)}{\psi'(1)}+o(1)\\
&= -\eta_2 \left(\phi_I + \phi_S-\xi\frac{\psi'(\theta)}{\psi'(1)} - \xi\zeta\right) - \beta\phi_I -\gamma(\phi_I-\xi\theta\zeta)  \\
&\qquad+ \xi \beta\phi_I\zeta +\xi \beta\phi_I\theta^2\frac{\psi''(\theta)}{\psi'(1)} - \xi\beta\phi_I\theta\left(\frac{\psi'(\theta)}{\psi'(1)}  + \theta\frac{\psi''(\theta)}{\psi'(1)}\right) -2\beta\theta\phi_I\frac{\psi'(\theta)}{\psi'(1)}+o(1)\\
&= -\eta_2 \left(\phi_I + \phi_S-\xi\frac{\psi'(\theta)}{\psi'(1)} - \xi\zeta\right) - \gamma(\phi_I-\xi\theta\zeta)  \\
&\qquad+ \xi \beta\phi_I\theta^2\frac{\psi''(\theta)}{\psi'(1)} - \xi\beta\phi_I\theta\left( \theta \frac{\psi''(\theta)}{\psi'(1)}\right) +o(1)
\end{align*}
Where in the last step we have used the fact that $\beta = o(1)$ and so anything which is bounded and multiplied by $\beta$ is also $o(1)$.   Continuing we have
\begin{align*}
\diff{}{t} \left(\phi_I +\phi_S - \xi \left[\theta \zeta + \theta^2\frac{\psi'(\theta)}{\psi'(1)}\right] \right)  &=-\eta_2 \left(\phi_I + \phi_S-\xi\frac{\psi'(\theta)}{\psi'(1)} - \xi\zeta\right) - \gamma(\phi_I-\xi\theta\zeta) +o(1) \\
 &=-\eta_2 \left(\phi_I + \phi_S-\xi\frac{\psi'(\theta)}{\psi'(1)} - \xi\zeta\right) - \gamma\left(\phi_I-\xi\theta\zeta+\phi_S - \xi\frac{\psi'(\theta)}{\psi'(1)}\right) +o(1) \\
&= - (\eta_2+\gamma) \left(\phi_I +\phi_S - \xi \left[\theta \zeta + \theta^2\frac{\psi'(\theta)}{\psi'(1)}\right] \right)  + o(1)
\end{align*}
[in the first step we used the fact that $\phi_S = \xi\psi'(\theta)/\psi'(1) + o(1)$].  So $\phi_I+\phi_S-\xi[\theta\zeta + \theta^2\psi(\theta)/\psi'(1)] = o(1)$.  Because $\theta=1+o(1)$ and $\phi_S=\xi\psi'(\theta)/\psi'(1)+o(1)$ it follows then that 
$\phi_I = \xi \theta \zeta + o(1)$.

\item $\dot{\theta} = -\tilde{\beta}\theta \zeta + \beta o(1)$.

This step is trivial.  Since $\phi_I = \xi\theta\zeta + o(1)$, we have 
\begin{align*}
\dot{\theta} &= -\beta \phi_I \\
&= - \beta \xi \theta \zeta + \beta o(1)\\
&= - \tilde{\beta} \theta \zeta + \beta o(1)
\end{align*}

\item $\zeta = 1 - \theta\frac{\psi'(\theta)}{\psi'(1)} + (\gamma/\tilde{\beta})\ln \theta +t o(1)$ 

We make the observation that $\dot{\zeta} = -\diff{}{t} [\theta\psi'(\theta)/\psi'(1)] - \gamma \zeta$.   
We have
\begin{align*}
\dot{\zeta} - \diff{}{t} \left[ 1 - \theta\frac{\psi'(\theta)}{\psi'(1)} + (\gamma/\tilde{\beta})\ln \theta\right] &= -\diff{}{t} \left[\theta\frac{\psi'(\theta)}{\psi'(1)}\right] - \gamma \zeta + \diff{}{t} \left[\theta\frac{\psi'(\theta)}{\psi'(1)}\right] - \frac{\gamma \dot{\theta}}{\tilde{\beta} \theta}\\
&= -\gamma \zeta - \frac{\gamma\dot{\theta}}{\tilde{\beta}\theta}
\end{align*}
From our previous result we have $\zeta = - [\dot{\theta}+\beta o(1)]/\tilde{\beta}\theta$.  Substituting this in we have
\begin{align*}
\dot{\zeta} - \diff{}{t} \left[ 1 - \theta\frac{\psi'(\theta)}{\psi'(1)} + (\gamma/\tilde{\beta})\ln \theta\right] &= - \frac{\gamma\dot{\theta}}{\tilde{\beta}\theta} - \frac{\gamma \beta o(1)}{\tilde{\beta}\theta} - \frac{\gamma \dot{\theta}}{\tilde{\beta}\theta}\\
= o(1)
\end{align*}
Our result follows immediately.
\end{enumerate}

Using the final two results, we have the equations of the MFSH model, with $\tilde{\beta}=\beta \xi$ playing the role of the transmission rate.  This ultimately leads to the MA model with $\hat{\beta} = \tilde{\beta}\ave{K}$ if the conditions of the previous section hold.
\small
\bibliographystyle{plain}
\bibliography{hierarchy}

\end{document}